%% file: paper.tex
\documentclass[manuscript,screen]{acmart}

\input{preamble.tex}

\acmJournal{TOSEM}

\begin{document}

\title{Leveraging Language Models for Log Statement Generation in Multilingual Scenarios: How Far Are We?}


\input{author_info.tex}

\input{section/abstract.tex}





\maketitle

\input{section/intro.tex}

\input{section/relatedwork.tex}

\input{section/experimentalsetup.tex}

\input{section/benchmark_construction.tex}
\input{section/results.tex}

\input{section/discussion.tex}

\input{section/threats.tex}
\input{section/conclusion.tex}
\input{section/data_availability.tex}

\begin{acks}
We gratefully acknowledge the financial support of: (1) JSPS for the KAKENHI grants (JP24K02921, JP25K03100, JP25K22845, JP26H02500); (2) Japan Science and Technology Agency (JST) as part of Adopting Sustainable Partnerships for Innovative Research Ecosystem (ASPIRE), Grant Number JPMJAP2415, and (3) the Inamori Research Institute for Science for supporting Yasutaka Kamei via the InaRIS Fellowship.
\end{acks}
  
\bibliographystyle{ACM-Reference-Format}
\bibliography{reference}

\end{document}

%% file: preamble.tex
\AtBeginDocument{%
  \providecommand\BibTeX{{%
    \normalfont B\kern-0.5em{\scshape i\kern-0.25em b}\kern-0.8em\TeX}}}

\usepackage{float}
\usepackage{hyperref}
\usepackage{booktabs}
\usepackage{algorithmic}
\usepackage{textcomp}
\usepackage{xcolor}
\usepackage{soul}
\usepackage{url}
\usepackage{xspace}
\usepackage{multirow}
\usepackage{enumitem}
\usepackage{colortbl}
\usepackage[skins,breakable]{tcolorbox}
\usepackage{framed}
\usepackage{hhline}
\usepackage{threeparttable}
\usepackage{url}
\usepackage{subcaption}
\usepackage{comment}
\usepackage{wrapfig}
\usepackage{subcaption}
\usepackage{caption}
\tcbuselibrary{raster,skins}

\definecolor{mygray}{gray}{0.9}
\definecolor{ForestGreen}{HTML}{288c66}
\definecolor{MyBlue}{HTML}{00008b}
\setulcolor{MyBlue}

\newcounter{examplecounter}[subsection]
\renewcommand{\theexamplecounter}{\arabic{section}.\arabic{subsection}.\arabic{examplecounter}}
\newlist{stepitemize}{itemize}{1}
\setlist[stepitemize,1]{leftmargin=1.45cm}

\newlist{phaseitemize}{itemize}{1}
\setlist[phaseitemize,1]{leftmargin=1.70cm}

\newcommand{\RQone}{What is the performance of log statement generation approaches in multilingual environments?}
\newcommand{\RQtwo}{What is the impact of training strategies on log statement generation in multilingual environments?}
\newcommand{\RQthree}{How do language-specific characteristics affect automated log statement generation in multilingual environments?}

\newlist{rqlist}{description}{1}
\setlist[rqlist]{%
  font=\bfseries,      
  style=nextline,      
  labelsep=0.6em,      
  leftmargin=!,        
  widest={(RQ3)},      
  itemsep=.6ex, topsep=.6ex
}

\newcommand{\Approach}{\textit{Approach:} }
\newcommand{\Results}{\textit{Results:} }


%% file: author_info.tex
\author{Kazuki Kusama}
\orcid{0009-0001-6623-3988}
\affiliation{%
\institution{Kyushu University}
\streetaddress{744 Motooka}
\city{Nishi-ku}
\state{Fukuoka}
\country{Japan}
\postcode{819-0395}
}
\email{kusama@posl.ait.kyushu-u.ac.jp}

\author{Honglin Shu}
\orcid{0009-0005-7311-7060}
\affiliation{%
  \institution{Kyushu University}
  \streetaddress{744 Motooka}
  \city{Nishi-ku}
  \state{Fukuoka}
  \country{Japan}
  \postcode{819-0395}
}
\email{shu.honglin.167@s.kyushu-u.ac.jp}

\author{Masanari Kondo}
\orcid{0000-0002-6317-7001}
\affiliation{%
  \institution{Kyushu University}
  \streetaddress{744 Motooka}
  \city{Nishi-ku}
  \state{Fukuoka}
  \country{Japan}
  \postcode{819-0395}
}
\email{kondo@ait.kyushu-u.ac.jp}

\author{Yasutaka Kamei}
\orcid{0000-0002-7058-1045}
\affiliation{%
  \institution{Kyushu University}
  \streetaddress{744 Motooka}
  \city{Nishi-ku}
  \state{Fukuoka}
  \country{Japan}
  \postcode{819-0395}
}
\email{kamei@ait.kyushu-u.ac.jp}

%% file: section/abstract.tex
\begin{abstract}

Log statements capture critical information for software maintenance activities such as testing, debugging, and failure analysis.
Because of this importance, developers must carefully design log statements, which requires significant effort.
To support developers, various end-to-end automated log statement generation approaches have been proposed, whereas these approaches have mainly been evaluated within a single programming language environment and their effectiveness in multilingual environments remains underexplored.
In this paper, we therefore comparatively evaluate three state-of-the-art log statement generation approaches and five large language models (LLMs) across multiple programming languages.
For this purpose, we constructed a multilingual benchmark comprising 150,000 instances across five programming languages.
Our empirical results demonstrate that UniLog, a state-of-the-art approach, achieves the best overall performance, maintaining high effectiveness even in multilingual environments.
We also observe substantial variance in the difficulty of log generation across languages: Python presents a greater challenge, whereas JavaScript yields comparatively better performance.
Detailed analysis reveals that these disparities stem from variations in log insertion distributions and language-specific logging idioms.
Our findings indicate that simply scaling model size or the volume of training data is insufficient for multilingual log generation; rather, designing approaches tailored to the specific characteristics of target languages is crucial.
These findings suggest that future automated logging techniques should explicitly account for language-specific logging characteristics to achieve robust performance in multilingual software development environments.
\end{abstract}

%% file: section/intro.tex
\section{Introduction} \label{sec:introduction}
Integrating log statements into code is a fundamental practice for capturing the dynamic internal behavior of software systems.
These log statements serve as an indispensable resource, driving critical maintenance activities such as testing, debugging, and failure analysis~\cite{Chen2021ACMCS,He2021ACMCS}.
To ensure that these log statements provide actionable insights while avoiding the system overhead of excessive logging and the information loss of insufficient logging, developers must carefully orchestrate their design by pinpointing optimal insertion locations, assigning appropriate severity levels, and formulating context-rich messages.
Because this design process inherently relies on the manual judgment of developers, it is notoriously challenging and imposes a substantial burden on developer time and effort~\cite{Yuan2012ICSE,Fu2014ICSE,Zhu2015ICSE,Li2021TSE}.

To alleviate the burden of manual log statement design, prior research has actively explored automated log statement generation~\cite{Guang2023TSE}.
Previously, this process was typically decomposed into discrete subtasks, such as identifying insertion locations, determining appropriate severity levels, and generating message content, and developed specialized automation techniques for each~\cite{Li2021ASE,Liqun2022ISSTA}.
Advances in language models have led researchers to propose end-to-end approaches that seamlessly integrate these subtasks to holistically automate log generation ~\cite{Mastropaolo2022ICSE,Xie2024ISSTA,Xu2024ICSE}.
For instance, UniLog~\cite{Xu2024ICSE} is an end-to-end approach for log statement generation that combines retrieval-based few-shot prompting with a warmup strategy to achieve strong performance.

Despite significant advances in automated log statement generation with language models, current evaluation approaches still suffer from three important limitations:

\smallskip{}
\noindent{}
\textbf{(I) The performance of log statement generation approaches in multilingual environments remains underexplored.}
Existing end-to-end log statement generation studies have been evaluated primarily on a single programming language, especially Java, and it remains unclear whether these approaches are equally effective in multilingual environments~\cite{Mastropaolo2022ICSE,Xie2024ISSTA,Xu2024ICSE}.
Modern software development commonly involves the use of multiple programming languages, such as Python and JavaScript~\cite{Li2022ESEC}.
This makes it necessary to move beyond single-language settings and evaluate the performance of log statement generation approaches across multiple programming languages.

\smallskip{}
\noindent{}
\textbf{(II) The impact of training strategies on multilingual log statement generation remains unclear.}
Existing log statement generation approaches adopt different training and adaptation strategies, such as fine-tuning pre-trained language models, retrieval-based prompting, and lightweight warmup.
In multilingual environments, it remains unclear whether models should be adapted separately for each programming language or trained jointly across multiple languages.
Language-specific adaptation may better capture local logging practices, vocabulary, and coding styles, whereas multilingual training may help models learn common patterns shared across languages.
Existing studies have not sufficiently investigated how such training strategies affect performance in multilingual log statement generation.
Without clarifying this point, determining how training data and model adaptation should be designed for practical multilingual log statement generation remains difficult.

\smallskip{}
\noindent{}
\textbf{(III) It remains unclear why automated log statement generation exhibits different performance across programming languages.}
Even if multilingual evaluation reveals performance differences across programming languages, such results alone do not explain why prediction is easier in some languages and more difficult in others.
Log statement generation involves multiple decisions, including where to insert logs, which log levels to assign, and what message content to generate.
These decisions may be influenced by language-specific factors, such as control structures, library usage, coding styles, and logging idioms.
However, existing studies have not sufficiently analyzed how such language-specific logging characteristics are related to performance differences across programming languages.
Without clarifying these factors, it remains difficult to explain cross-language performance gaps and to derive design guidelines for log statement generation approaches suitable for multilingual environments.

To address these three limitations, this study empirically investigates end-to-end log statement generation in multilingual environments.
Specifically, we construct a comprehensive multilingual benchmark from public GitHub repositories across five programming languages: Java, Python, JavaScript, TypeScript, and C\#.
For each language, the benchmark contains 24,000 training instances, 3,000 validation instances, and 3,000 test instances, resulting in a total of 150,000 instances.
Using this benchmark, we comparatively evaluate the performance of three representative existing approaches and five large language models (LLMs) under unified experimental conditions, including the same target languages, data construction criteria, and evaluation metrics.
We also analyze the effects of different training strategies and examine language-specific logging characteristics, including log position, log level, and log message, thereby clarifying cross-language differences and the factors underlying them.
We define the following three research questions (RQs):
\begin{rqlist}
    \item [(RQ1) \RQone]
    \Approach{}
    We evaluate three representative existing approaches, LANCE~\cite{Mastropaolo2022ICSE}, FastLog~\cite{Xie2024ISSTA}, and UniLog~\cite{Xu2024ICSE}.
    Since general-purpose LLMs have shown promising performance in log statement generation, we also evaluate five LLMs in multilingual environments: Llama3, Qwen2.5-Coder, Mistral, DeepSeek-V3, and GPT-4.1 mini.\\
    \Results
    UniLog achieved the highest performance among the existing approaches in multilingual environments and perfectly predicted log statements with an average of 20.35\% across all languages.
    Among the evaluated LLMs, DeepSeek-V3 achieved the highest performance and perfectly predicted log statements with an average of 10.96\% across all languages.
    In addition, JavaScript showed the highest overall performance, whereas Python showed the lowest overall performance, which indicates that prediction difficulty differs across languages.
    \item [(RQ2) \RQtwo]
    \Approach{}
    We evaluate multiple training strategies, including monolingual and multilingual training, warmup-based adaptation, and training strategies used in representative end-to-end approaches.\\
    \Results
    Single-language training showed overall better performance than mixed-language training.
    UniLog also showed strong performance, although it requires less training data.

    \item [(RQ3) \RQthree]
    \Approach{}
    We examine insertion locations, log levels, and message content across programming languages to clarify the factors underlying cross-language performance differences.\\
    \Results
    Language-specific logging characteristics, such as insertion locations, log levels, and message content, differ across programming languages.
    For example, log statements inside loops, which are difficult to predict, are more common in Python and less common in JavaScript.
    This tendency partially explains the performance differences across languages.
\end{rqlist}

The main contributions of this study are as follows:
\begin{itemize}
    \item We release a large-scale multilingual benchmark~\footnote{\url{https://doi.org/10.5281/zenodo.20279312}} for end-to-end log statement generation, collected from public GitHub repositories and spanning five programming languages: Java, Python, JavaScript, TypeScript, and C\#.
    For each language, the benchmark comprises 24,000 training, 3,000 validation, and 3,000 test instances, amounting to 150,000 instances in total.
    \item We conduct an empirical study to evaluate the multilingual adaptability of current log generation approaches. 
    Specifically, we benchmark three representative existing approaches (LANCE, FastLog, and UniLog) against five state-of-the-art LLMs (Llama3, Qwen2.5-Coder, Mistral, DeepSeek-V3, and GPT-4.1 mini) using a unified dataset and standardized evaluation metrics.
    \item To ensure reproducibility and transparency, we release the dataset, experimental code, and results used in this study.
    This enables the research community to verify the findings of this study and promote further research.
\end{itemize}

%% file: section/relatedwork.tex
\section{Related Work} \label{sec:relatedwork}
This section reviews studies on automated log statement generation and studies on multilingual research in software engineering.

\subsection{Log Statement Generation} \label{subsec:logstatementgeneration}

Previous log statement generation research first focused on automating the subtasks~\cite{Li2018ESE,Li2021ASE,Li2021ICSE,Ding2022SANER,Ding2023ACM}.
For example, Li et al.~\cite{Li2021ASE} proposed a deep learning framework that estimates log insertion locations at the code block level by incorporating syntactic and semantic information from source code.
For log level prediction, Li et al.~\cite{Li2021ICSE} proposed DeepLV, which models log level recommendation as an ordinal classification task using syntactic context and log message features extracted from source code.
Ding et al.~\cite{Ding2022SANER} proposed LoGenText, which generates natural language log messages based on neural machine translation.

Around 2022, research more focused on end-to-end log statement generation using language models~\cite{Mastropaolo2022ICSE,Xie2024ISSTA,Xu2024ICSE,Li2025ASE}. 
Mastropaolo et al.~\cite{Mastropaolo2022ICSE} proposed LANCE, which is the first pre-trained language model (PLM)-based end-to-end approach built on T5 with fine-tuning.
LEONID, which extends LANCE, has also been proposed~\cite{Mastropaolo2024JSS}.
LEONID integrates deep learning and information retrieval, which is abbreviated as IR, to improve accuracy, but it involves a substantial increase in computational cost while achieving only limited performance improvement.
Xie et al.~\cite{Xie2024ISSTA} proposed FastLog, which is another PLM-based end-to-end log statement generation approach.
FastLog introduces a design that splits the input code and predicts insertion locations at the token level in order to improve the accuracy of log insertion location prediction.
Xu et al.~\cite{Xu2024ICSE} proposed UniLog, which is an LLM-based end-to-end log statement generation approach that combines few-shot prompting with a warmup strategy, which is a lightweight fine-tuning approach.
UniLog reported high performance in log insertion location and log level prediction and demonstrated the effectiveness of retrieval-based demonstrations and lightweight model adaptation.
Li et al.~\cite{Li2025ASE} proposed LOGIMPROVER, which is an LLM-based end-to-end logging automation framework for logging quality improvement.
LOGIMPROVER combines candidate identification, contextualized refinement, false positive pruning, and holistic logging patch generation to improve logging quality.
Their results demonstrated the effectiveness of this end-to-end framework in large-scale codebases.
In addition to these end-to-end approaches, Li et al.~\cite{Li2024TSE} conducted the first empirical study on the effectiveness of LLMs for automated logging generation.
They evaluated multiple LLMs on both code collected from open-source projects and semantically equivalent transformed variants, and showed that, although LLMs are promising for logging generation, their performance and generalization capability remain limited.

While there are various automated log statement generation approaches, prior studies have mainly evaluated these approaches in limited language settings, primarily on Java and, in some cases, Go.
As a result, empirical evidence remains limited regarding how current log statement generation approaches perform across multiple programming languages.


\subsection{Multilingual Software Engineering Research} \label{subsec:multilingual}

In real software development, it is common to use multiple programming languages together \cite{Li2022ESEC,Li2023TOSEM}.
Li et al. \cite{Li2023TOSEM} characterized how open-source multilingual systems are constructed and how programming languages are selected in such systems, providing further evidence that multilingual software development is a common and practically important setting.
Prior work has also shown that multilingual software development involves diverse issues and challenges arising from cross-language interfacing, data handling, and the complexity of language-specific features \cite{Yang2023ICSE}.
In addition, Yang et al. \cite{Yang2022FSE} argued that analyzing multilingual code holistically is important, while fully language-agnostic designs may face practicality challenges.
To demonstrate the practical usefulness of approaches in software engineering, it is important not only to confirm effectiveness on a single language but also to evaluate generality across multiple languages.
From this perspective, benchmark construction and empirical evaluation for software engineering tasks in multilingual environments have been actively studied \cite{Wang2025TOSEM,Yang2024FSE,shu2025EMSE,zan2025NeurIPS,Shirai2026MSR}.

Zan et al.~\cite{zan2025NeurIPS} proposed SWE-bench-Multi, which extends SWE-bench, a benchmark widely used in the context of automated program repair, to eight languages, and they provided a benchmark for multilingual automated program repair.
Their empirical study showed that existing program repair approaches that were effective on SWE-bench often generalize poorly to languages other than Python.
As possible reasons, they reported that existing approaches are mainly optimized for Python and that task difficulty differs across languages.

Wang et al.~\cite{Wang2025TOSEM} empirically evaluated the performance of PLMs and LLMs across seven programming languages in the context of automated vulnerability repair.
They found that repair success rates vary substantially across programming languages, with the highest success rate of 31.59\% on Go, whereas the success rate on C++ was only 6.73\%.
Yang et al.~\cite{Yang2024FSE} addressed multilingual bug detection and localization by proposing a deep learning-based technique for detecting and localizing multilingual bugs.
Their work highlighted that bug detection and localization support has long been centered on single-language software despite the prevalence of multilingual systems, and demonstrated the importance of modeling cross-language structures in multilingual settings.
Shu et al.~\cite{shu2025EMSE} evaluated the performance of PLMs and LLMs across seven programming languages in the context of automated vulnerability detection.
They reported that detection performance also varies across programming languages, with the highest accuracy of 80.82\% on Go, whereas the lowest accuracy of 66.26\% on Python, and they confirmed clear cross-language differences in multilingual environments.

These studies indicate that multilingual evaluation is important in software engineering and support the necessity of verifying the effectiveness of approaches across different programming languages.
As discussed, existing log statement generation research has mainly evaluated approaches in single-language settings.
Hence, in this paper, we aim to construct a multilingual benchmark for log statement generation and conduct a large-scale empirical study to evaluate the performance of log statement generation approaches. To deeply understand the log statement generation on multilingual environments, we investigate not only the performance, but also the impact of LLM training strategies and language-specific logging characteristics across multiple programming languages.

%% file: section/experimentalsetup.tex
\section{EXPERIMENTAL SETUP} \label{sec:experimentalsetup}
\input{figures/experimental_setup/overview_figure.tex}
This section describes the experimental design of this study, as shown in Figure~\ref{fig:overview}.
We first provide an overview of the evaluated log statement generation approaches and training strategies.
We then describe the implementation details and evaluation metrics used in this study.
We evaluate the performance of log statement generation approaches, including existing end-to-end approaches and LLMs, in multilingual environments (RQ1).
Next, we investigate the effect of different training strategies on multilingual log statement generation performance (RQ2).
Finally, we analyze language-specific characteristics of log statements to identify factors that affect prediction difficulty across programming languages (RQ3).

\subsection{Experimental Techniques and Settings} 
\label{sec:experimentalsetup_models}
In this study, we evaluate the performance of existing end-to-end approaches and LLMs, as well as the impact of training strategies, on multilingual log statement generation.
This section describes the approaches and settings compared in this study.

\subsubsection{Existing End-to-End Log Statement Generation Approaches} \label{sec:experimentalsetup_baselines}
To evaluate how effectively existing end-to-end log statement generation approaches work in multilingual environments, we conducted a literature review of related papers published in the software engineering community and selected baselines.
Specifically, we selected the following three approaches, which are widely used as baselines in log statement generation research~\cite{Zhong2025TOSEM,Shu2025TOSEM}:

\begin{itemize}
    \item \textbf{LANCE}~\cite{Mastropaolo2022ICSE} is the first end-to-end approach leveraging a pre-trained language model (PLM) for automated log statement insertion.
    Built upon the T5 architecture, LANCE first acquires a general understanding of Java syntax and typical logging locations through a dedicated pre-training phase. It is subsequently fine-tuned to accurately synthesize and inject contextually appropriate log statements directly into the source code.

    \item \textbf{FastLog}~\cite{Xie2024ISSTA} extends the foundational concepts of LANCE~\cite{Mastropaolo2022ICSE} by introducing a decoupled, two-stage PLM architecture.
    In the first stage, the model explicitly predicts optimal log insertion locations; in the subsequent stage, it synthesizes the corresponding log statements for those specific sites. 
    This architectural refinement enables FastLog to achieve both higher accuracy and significantly faster generation speeds compared to its predecessor.

    \item \textbf{UniLog}~\cite{Xu2024ICSE} represents an LLM-driven framework for end-to-end log statement generation.
    UniLog incorporates lightweight parameter tuning, which is referred to as warmup, and retrieval-based few-shot prompting to generate log insertion locations and log statements. 
\end{itemize}

\subsubsection{Evaluated LLMs} \label{sec:experimentalsetup_llms}
We studied five representative LLMs.
These models have been widely adopted in the literature on various software engineering tasks~\cite{Zhong2025TOSEM,Kazuki2025ESEM,shu2025EMSE,Acharya2025EASE,zan2025NeurIPS,Wang2025SC}.
In this study, we categorize them into two types based on how they are used, which we refer to as \textit{self-hosted} and \textit{API-based}.
\textit{Self-hosted} category refers to LLMs that can be deployed locally within our own computational infrastructure in the experimental environment of this study (Section~\ref{sec:experimentalsetup_implementation}), whereas \textit{API-based} refers to LLMs that are accessed through external service APIs. 
We strategically selected three \textit{self-hosted} LLMs with parameter counts ranging from 7B to 8B. This choice is motivated by deployment constraints: many researchers lack the high-end computational infrastructure necessary to deploy or instruction-tune significantly larger models.

The details of these LLMs are as follows:
\begin{itemize}
    \item \textbf{Llama3 (\textit{self-hosted})}~\cite{llama3} is a family of instruction-tuned LLMs released by Meta and designed for tasks that require multilingual dialogue, coding, and long-context processing.
    It has shown strong performance on a variety of tasks, including mathematical reasoning and code generation.
    In this study, we use \texttt{Llama3.1-8B-Instruct}.
    \item \textbf{Qwen2.5-Coder (\textit{self-hosted})}~\cite{qwen25coder} is a code-specialized LLM in the Qwen2.5~\cite{qwen25} family.
    It substantially improves code-related capabilities while maintaining the strengths of Qwen2.5 in mathematics and general abilities.
    In this study, we use \texttt{Qwen2.5-Coder-7B}.
    \item \textbf{Mistral (\textit{self-hosted})}~\cite{mistral} is an LLM released by Mistral AI and targets a wide range of tasks, including dialogue generation, reasoning, and code generation.
    Mistral is known for its efficient performance, which is supported by fast inference speed and broad context handling.
    In this study, we use \texttt{Mistral-7B-Instruct-v0.3}.
    \item \textbf{GPT-4.1 mini (\textit{API-based})}~\cite{GPT4-mini-report} is a small LLM provided by OpenAI that is fast and low-cost and is suitable for tasks that require instruction following.
    It substantially outperforms GPT-4o while reducing latency and cost.
    \item \textbf{DeepSeek-V3 (\textit{API-based})}~\cite{deepseek-v3} is a Mixture-of-Experts (MoE) LLM that was pre-trained on more than 14T tokens and supports a 128k context window.
    It has been reported to achieve state-of-the-art performance among open-source models and to outperform strong models such as GPT-4o~\cite{gpt4o} and Claude3.5~\cite{claude35}.
\end{itemize}

\subsubsection{Strategies for LoRA Fine-tuning} \label{sec:experimentalsetup_lora_strategies}
In this study, we investigate the effect of training strategies on multilingual log statement generation.
Specifically, we compare training strategies for parameter-efficient fine-tuning (PEFT) with LoRA~\cite{Edward2021arXiv}.
LoRA is an approach that adapts a foundation model to a task with a small number of additional parameters by freezing the weights of the base model and learning only low-rank matrices.
Prior work has reported that LoRA achieves better performance on log statement generation tasks than other PEFT approaches, such as Prompt tuning and Prefix tuning~\cite{Zhong2025TOSEM}.
In this study, we compare the following two strategies as learning units for LoRA.
\begin{itemize}
    \item \textbf{Monolingual LoRA (Mono-LoRA)}: We perform LoRA fine-tuning independently for each language.
    In this setting, language-specific adaptation parameters are learned for each language in order to acquire log statement generation capability specialized for that language.
    This allows us to evaluate the effect of locally incorporating language-specific vocabulary, expressions, and logging practices.

    \item \textbf{Multilingual LoRA (Multi-LoRA)}: We integrate the training data of the five languages and train a single LoRA.
    After training, the same adaptation parameters are applied to all languages during inference.
    In this setting, we evaluate whether cross-language log statement generation capability can be acquired by capturing code patterns and log insertion strategies that are shared across multiple languages.
\end{itemize}

\subsection{Implementation} \label{sec:experimentalsetup_implementation}
This subsection describes the implementation details, task design, and prompt design for the existing end-to-end log statement generation approaches and LLMs.

\underline{\textit{LANCE.}}
\input{figures/experimental_setup/LANCE_figure.tex}
Figure~\ref{fig:LANCE} illustrates the input and output formats of LANCE for both fine-tuning and inference. 
LANCE receives a source method with the target log statement removed as input and generates the complete method with the predicted log statement inserted.
Since LANCE originally targets Java, we adapted the framework to support multilingual scenarios by modifying the backbone model and training strategy, as described below.
\begin{itemize}
    \item \textbf{CodeT5+ Backbone:} We replace the general-text T5 backbone with CodeT5+. Because CodeT5+ is explicitly pre-trained on a massive multilingual code corpus, unlike T5, which primarily targets natural language, this substitution eliminates the prohibitive cost of collecting and pre-training on language-specific code corpora.
    \item \textbf{Omission of Pre-training:} We bypass the resource-intensive pre-training phase of LANCE and rely exclusively on fine-tuning. This decision is twofold: (1) the CodeT5+ backbone already encapsulates prior knowledge across various programming languages, and (2) prior ablation studies~\cite{Mastropaolo2022ICSE} indicate that this specific pre-training step yields only marginal performance gains.
\end{itemize}

To accommodate these changes, we aligned our hyperparameters with standard CodeT5+ configurations, specifically setting the learning rate to 2e-5 and the batch size to 16. 
To empirically validate these adjustments, we fine-tuned CodeT5+ on the original LANCE dataset and successfully reproduced the reported performance of the paper within a margin of three points.

\underline{\textit{FastLog.}}
\input{figures/experimental_setup/FastLog_figure.tex}
Figure~\ref{fig:FastLog} illustrates the input and output formats of FastLog during both fine-tuning and inference.
FastLog automates log statement insertion via a sequential, two-stage pipeline.
In Stage-1, the model performs a token classification task over each token in the input method, predicting whether a log statement should be inserted immediately after that token; during inference, the token with the highest probability of being classified as \texttt{1} is selected as the insertion location.
In Stage-2, a \texttt{<mask>} token is inserted as a placeholder immediately after the predicted insertion location, and the log statement body is generated in a sequence-to-sequence manner to fill the \texttt{<mask>}. Finally, the \texttt{<mask>} is replaced with the generated log statement, yielding the complete method with the inserted log statement without regenerating the surrounding non-log code.

In this study, we leveraged the official FastLog replication package to fine-tune two distinct PLBART-based~\cite{PLABART} models: a token classification model for Stage-1 and a log generation model for Stage-2. 
Adhering closely to the original configurations of the authors, we applied a learning rate of 2e-5 and a batch size of 8. The Stage-1 model was trained for 10 epochs, while the Stage-2 model underwent 30 epochs of fine-tuning.

\underline{\textit{UniLog.}}
\input{figures/experimental_setup/LLM_figure.tex}
Figure~\ref{fig:LLM} illustrates the input and output formats of UniLog, an end-to-end LLM-based approach for log statement generation by lightweight parameter tuning (Warmup) and retrieval-based few-shot prompting. 
Given a source method with the target log statement removed, UniLog simultaneously predicts the optimal insertion location (line ID) and generates the corresponding log statement (level and message).

Because the original backbone of UniLog, Codex, is deprecated, we use GPT-4.1 mini as the backbone in our reimplementation due to its favorable balance between cost-efficiency and performance.
The implementation of UniLog centers on two primary mechanisms: \textbf{(1) Retrieval-Augmented Prompting.} For both warmup and inference, prompts consist of a task instruction, the target method (Query), and the top-five most similar training examples retrieved via cosine similarity. To build spatial awareness, code lines in the prompt are explicitly annotated with <line\#> tags so that the model can link code context directly to the predicted line ID. \textbf{(2) Warmup.} Prior to inference, the model undergoes parameter tuning using 500 randomly selected validation queries. This allows the LLM to adapt specifically to the logging patterns of the target dataset. During inference, the same prompt construction is applied to the test data to evaluate the generated line IDs and log statements.

\underline{\textit{General-purpose LLMs.}}
To formulate the task and construct the prompts for the evaluated LLMs, we mirror the architecture established by UniLog, since it is an LLM-based baseline in our study and using the same task setting enables a fairer comparison among LLM-based approaches.
Specifically, the model receives a source method with the target log statement removed as input, and it outputs both the predicted insertion location and the complete log statement (comprising the severity level and message content).

We employ retrieval-based few-shot prompting, a strategy demonstrated by Zhong et al.~\cite{Zhong2025TOSEM} to consistently outperform alternative techniques such as Chain-of-Thought. 
Adhering to the prompt structure of UniLog, which encompasses a task instruction, an Example with its corresponding Label, and a Query, we specifically utilize a 1-shot setting~\cite{Zhong2025TOSEM}. 
For each test method (the Query), we retrieve the single most relevant training instance based on cosine similarity to serve as the Example. To guarantee reproducibility across generations, the decoding temperature is strictly set to 0.

To investigate the impact of training strategies on LLM performance in RQ2, we fine-tune the models using Low-Rank Adaptation (LoRA). 
The input-output formulation and prompt template remain identical to those used during inference, with queries sourced from the validation dataset. 
For the LoRA hyperparameters, we configure the rank $r$ to 16, the scaling coefficient $\alpha$ to 32, the learning rate to 1e-4, and the batch size to 64.

\textbf{Experiment Environment.}
GPT-4.1 mini was used through the OpenAI API.
Specifically, we used gpt-4.1-mini-2025-04-14 as the experimental model version of GPT-4.1 mini.
DeepSeek-V3 was used through the DeepSeek API.
All open-source LLMs were downloaded from Hugging Face, specifically Llama3, Qwen2.5-Coder, and Mistral.
For all LLMs, experiments were conducted on a machine equipped with 251GiB RAM, Ubuntu 24.04.3 LTS, and an RTX 6000 Ada Generation GPU with 48GB, with an Intel(R) Xeon(R) w7-2595X 26-core processor.

\subsection{Evaluation Metrics} \label{sec:experimentalsetup_metrics}
To evaluate the performance of automated log statement generation, we adopt evaluation metrics that are commonly used in end-to-end log statement generation research~\cite{Xie2024ISSTA,Xu2024ICSE}.
Specifically, we evaluate the correctness of the insertion location, log level, and message content of each log statement using the following measures:

\begin{itemize}
\item \textbf{Position Accuracy:} The proportion of cases in which the predicted insertion location exactly matches the ground truth.
\item \textbf{Level Accuracy:} The proportion of cases in which the predicted log level matches the ground truth.
\item \textbf{Message Accuracy:} The proportion of cases in which the generated message exactly matches the ground truth message.
\item \textbf{All Accuracy:} The proportion of cases in which the insertion location, log level, and message all simultaneously match the ground truth.
\end{itemize}

Because Message Accuracy is based on exact matching, it does not evaluate cases in which the predicted message and the reference message convey the same meaning but differ in expression. 
To evaluate how appropriately log messages are predicted, we additionally use the following text similarity metrics:

\begin{itemize}
    \item \textbf{BLEU}~\cite{BLEU} is an automatic evaluation metric that is widely used in machine translation and measures n-gram overlap between a generated sentence and a reference sentence.
    In this study, we report sentence-level BLEU with smoothing on a scale from 0 to 100.
    Following the standard definition, BLEU integrates 1 to 4 gram precision with uniform weights of 0.25 each and includes a brevity penalty.
    
    \item \textbf{ROUGE}~\cite{ROUGE} is an automatic evaluation metric that is widely used in summarization and measures lexical overlap between a generated sentence and a reference sentence.
    In this study, we report sentence-level ROUGE-L based on the longest common subsequence, on a scale from 0 to 100.
    ROUGE-L evaluates overlap while preserving order and provides a perspective different from that of BLEU.
\end{itemize}

These n-gram-based metrics have also been widely used in prior work to measure the similarity between generated log messages and target messages~\cite{Xie2024ISSTA,Shu2025TOSEM}.

%% file: figures/experimental_setup/overview_figure.tex
\begin{figure*}[t]
  \centering
  \includegraphics[width=0.9\textwidth]{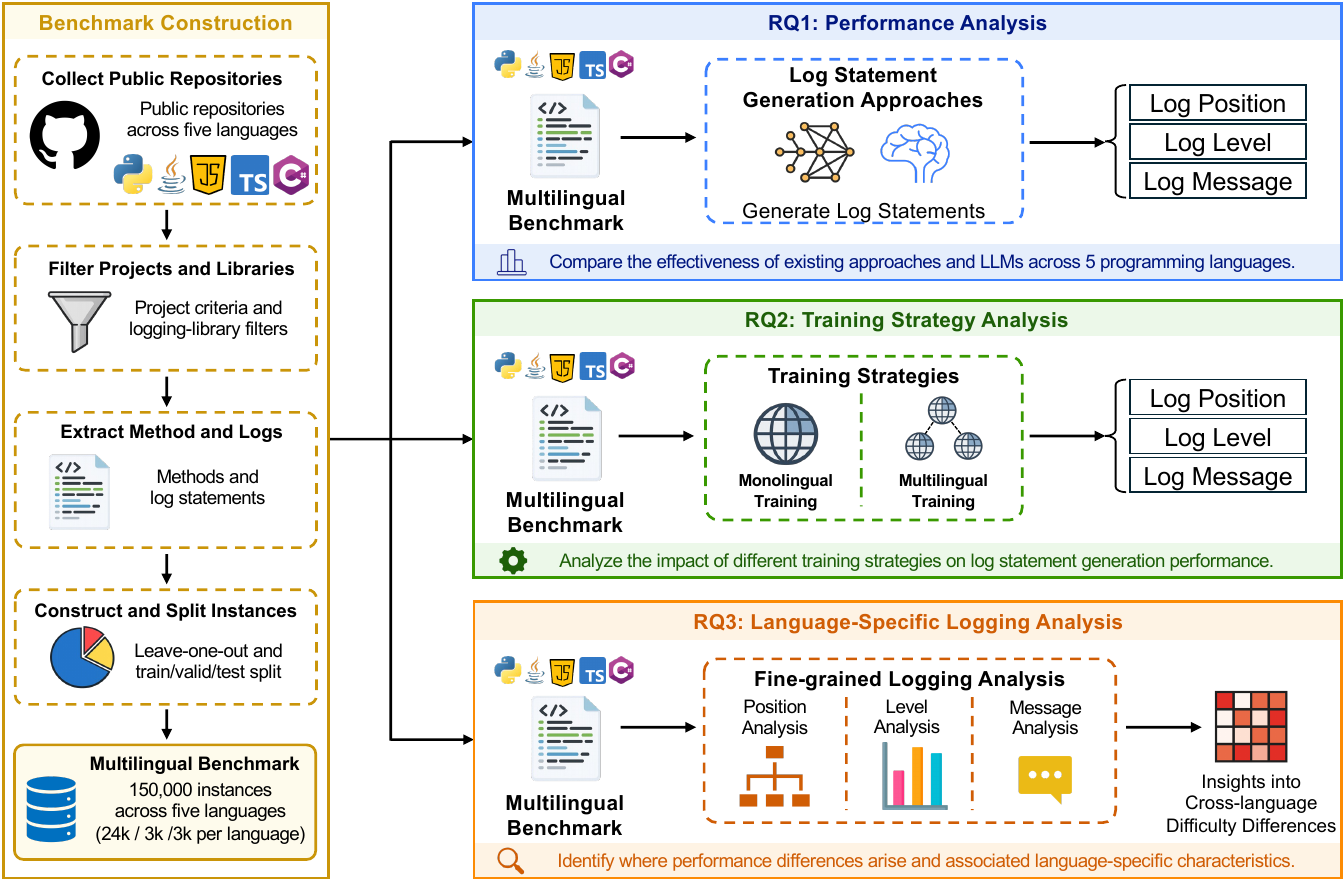}
  \caption{The overview of the experimental design.}
  \label{fig:overview}
\end{figure*}

%% file: figures/experimental_setup/LANCE_figure.tex
\begin{figure*}[t]
  \centering
  \includegraphics[width=.9\textwidth]{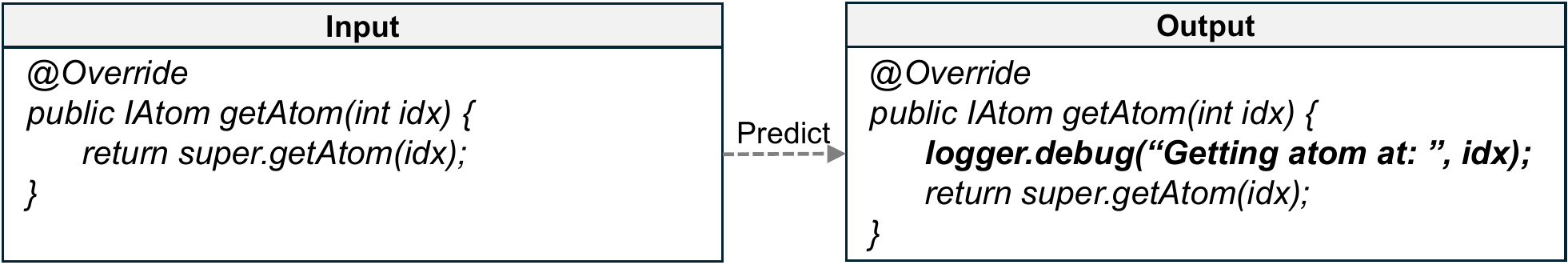}
  \caption{Input-output design of LANCE for fine-tuning and inference.}
  \label{fig:LANCE}
\end{figure*}

%% file: figures/experimental_setup/FastLog_figure.tex
\begin{figure*}[t]
  \centering
  \includegraphics[width=.9\textwidth]{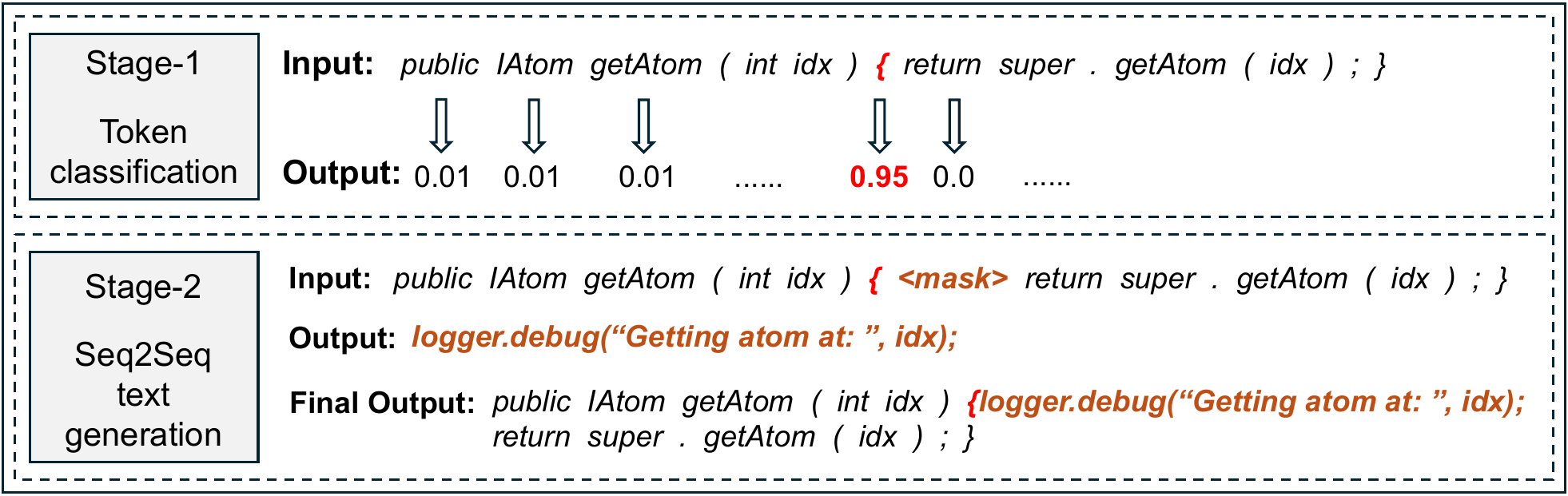}
  \caption{Input-output design of FastLog for fine-tuning and inference.}
  \label{fig:FastLog}
\end{figure*}

%% file: figures/experimental_setup/LLM_figure.tex
\begin{figure*}[t]
  \centering
  \includegraphics[width=.9\textwidth]{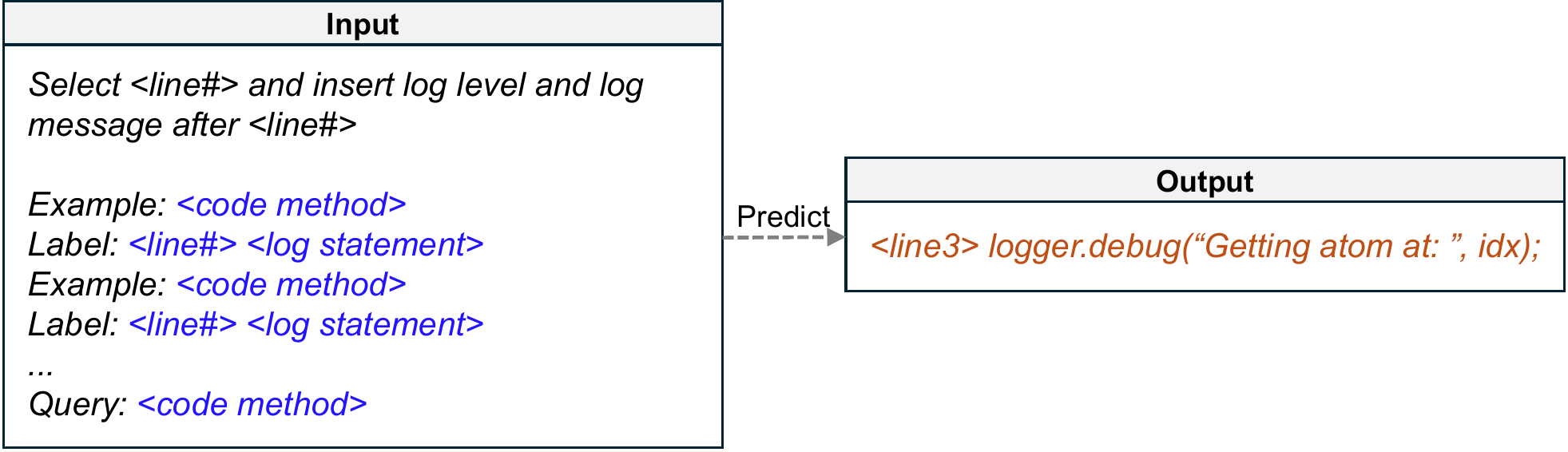}
  \caption{Input-output design of UniLog and LLMs for warmup and inference.}
  \label{fig:LLM}
\end{figure*}

%% file: section/benchmark_construction.tex
\section{Benchmark Construction} \label{sec:benchmarkconstruction}
To construct our multilingual dataset, we mined open-source repositories hosted on GitHub across five prominent programming languages: Java, Python, JavaScript, TypeScript, and C\#. 
We selected these five languages because they were the top five most commonly used programming languages on GitHub in 2025, according to the 2025 GitHub Octoverse report.~\footnote{https://github.blog/news-insights/octoverse/octoverse-a-new-developer-joins-github-every-second-as-ai-leads-typescript-to-1/}

\underline{\textit{Project Mining.}}
We leveraged the repository mining tool provided by~\citet{dabic2021MSR} to systematically extract candidate projects across the five target languages. To ensure dataset quality and align with established methodologies~\cite{Mastropaolo2022ICSE}, we filtered these repositories based on the following criteria:
\begin{itemize}
    \item Having at least 500 commits, 10 contributors, and 10 stars, to avoid toy or personal projects.
    \item Not being forked to reduce the chance of mining duplicated code.
\end{itemize}
Following this rigorous filtering phase, we retained 9,018 candidate projects for Java, 17,239 for Python, 12,728 for JavaScript, 11,581 for TypeScript, and 3,750 for C\#.

\underline{\textit{Logging Dependency Filtering.}}
To ensure dataset consistency, we restricted our selection to projects utilizing representative logging libraries or standard logging mechanisms specific to each language. 
Because substantial variations in logging APIs and severity schemes can destabilize log extraction and level normalization, filtering out non-standard implementations is crucial to prevent heterogeneous training targets. 
Consequently, we limited our dataset to projects explicitly relying on Log4j~\cite{log4j} for Java, the standard logging module~\cite{log_library} for Python, Winston~\cite{winston} or Pino~\cite{pino} for JavaScript and TypeScript, and NLog~\cite{nlog}, log4net~\cite{log4net}, or Serilog~\cite{serilog} for C\#. 
We identified these dependencies primarily by analyzing language-specific build configuration files (e.g., \texttt{pom.xml} for Java, \texttt{package.json} for JavaScript and TypeScript, and \texttt{.csproj} for C\#), supplementing this process when necessary by scanning source code for explicit \texttt{import} statements in Java and Python, \texttt{require} calls in JavaScript and TypeScript, and \texttt{using} directives in C\#.

\underline{\textit{Method Extraction and Filtering.}}
Log statement generation is typically formulated as a method-level task, where the model generates log statements for a given method~\cite{Mastropaolo2022ICSE,Xie2024ISSTA,Xu2024ICSE}.
We systematically extracted individual methods from the curated projects.
In alignment with established practices~\cite{Mastropaolo2022ICSE}, we retained only methods that satisfy \#tokens $\le$ 512 and \#tokens $>$ 10, where \#tokens denotes the number of tokens after excluding comments.
This constraint is crucial for mitigating the computational overhead associated with LLM training and inference.
To eliminate the risk of data leakage, we computed the hash value of each method and removed duplicate methods with identical hashes.

\underline{\textit{Log Statement Identification.}}
\input{table/experimentalsetup/dataset_stats.tex}
To identify log statements within the filtered methods, we implemented a pattern matching approach based on pairs of logger identifiers and log level names. 
Specifically, we identified method calls as log statements when log-level method names (e.g., \texttt{info}, \texttt{debug}, and \texttt{error}) were invoked on frequently used logger identifiers (e.g., \texttt{log}, \texttt{logger}, and \texttt{logging}).
To reduce missed extractions caused by notational variation across languages and libraries, we also normalized differences in letter case and synonymous expressions, such as \texttt{warning}/\texttt{warn} and \texttt{information}/\texttt{info}, treating them as equivalent categories. 
The \# Methods column in Table~\ref{tab:full_dataset_stats} reports the number of methods obtained for each language.

\underline{\textit{Code Formatting.}}
To reduce stylistic inconsistency and ensure data quality, we reformatted the extracted methods using automatic code formatting, following prior work~\cite{Xu2024ICSE}.
While the prior work applied Google Java Format to Java code, we extended this preprocessing step to a multilingual setting.
Specifically, we applied Google Java Format~\cite{google-java-format} to Java, and for the other languages, we selected widely used formatting tools with strong adoption in practice: Black~\cite{black} for Python, Prettier~\cite{prettier} for JavaScript and TypeScript, and dotnet format~\cite{dotnet-format} for C\#.

\underline{\textit{Instance Construction.}}
We generated our final datasets by converting the formatted methods into discrete data instances.
Building upon established methodologies from LANCE~\cite{Mastropaolo2022ICSE} and UniLog~\cite{Xu2024ICSE}, we structure each instance as an input-output pair comprising a source method and a single target log statement. 
For methods containing multiple log statements, we employ a leave-one-out extraction strategy. 
Specifically, only the focal target log statement is removed, while all other log statements remain intact to provide vital contextual information for the model.
Figure~\ref{fig:leave-one-out} illustrates this leave-one-out extraction process.
In this example, the original method contains two log statements; therefore, two instances are generated by removing one log statement at a time.
In each instance, the removed log statement serves as the expected output, while the remaining log statement is kept in the input method. 
By iteratively repeating this process, a single source method containing $n$ log statements generates exactly $n$ distinct instances.
The final yield of instances per language is detailed in the \# Instances column of Table~\ref{tab:full_dataset_stats}.
\input{figures/experimental_setup/leave-one-out_figure.tex}

\underline{\textit{Dataset Split.}}
Following prior work~\cite{Zhong2025TOSEM}, we partitioned the final dataset into training, validation, and test sets using a target ratio of 8:1:1.
The exact number of instances per set for each language is detailed in the Train, Valid, and Test columns of Table~\ref{tab:full_dataset_stats}.
To prevent data leakage, where highly similar code snippets from the same file inadvertently appear across multiple splits, we enforced a strict file-level splitting strategy.
This guarantees that all instances originating from a single file are assigned exclusively to one of the three sets.
To avoid confounding the effect of programming language with differences in data volume, we randomly sampled exactly 24,000 training, 3,000 validation, and 3,000 test instances per language.
Although the natural distribution of programming languages in real-world repositories is inherently imbalanced, we maintained a unified instance distribution across the five languages to enable a fair comparison of multilingual performance.
In particular, this design helps disentangle whether performance differences arise from language-specific characteristics or simply from differences in the number of fine-tuning instances available for each language.
This yielded a perfectly balanced multilingual benchmark comprising 30,000 instances per language (150,000 total instances).
To ensure reproducibility and facilitate future research, both the full dataset and this balanced experimental subset are publicly available in our replication package.

\underline{\textit{Log Level Normalization.}}
Table~\ref{tab:dataset_level} shows the distribution of log levels in the dataset for each language. Following prior work~\cite{Shu2025TOSEM}, we identified representative level expressions, such as \texttt{info}, \texttt{fatal}, and \texttt{debug}, using regular expressions, normalized notational variants that are semantically equivalent (e.g., synonyms and case differences), and mapped them to the following six standardized levels: \textit{Critical/Fatal}, \textit{Error}, \textit{Warn}, \textit{Info}, \textit{Debug}, and \textit{Trace/Verbose}.
\textit{Critical/Fatal} represents the most severe situations, capturing events that may directly lead to system shutdown or unrecoverable failures. 
\textit{Error} denotes the next severity tier, recording problems that may cause system failures or prevent continued processing. 
\textit{Warn} captures signs of potential degradation — such as unexpected states or inconsistencies — that do not immediately halt processing but may escalate to \textit{Error}. 
\textit{Info} records information pertaining to system operating status, such as progress updates and state transitions. 
\textit{Debug} captures diagnostic information used to identify root causes during development and troubleshooting. 
\textit{Trace/Verbose} is the most granular level, recording fine-grained details that enable tracing of program execution flow, such as method calls, branch traversal, and internal state changes.
As shown in Table~\ref{tab:dataset_level}, \textit{Error}, \textit{Info}, and \textit{Debug} account for the majority of log statements across all languages.

\input{table/experimentalsetup/dataset_statistics.tex}

%% file: table/experimentalsetup/dataset_stats.tex
\begin{table}[t]
\centering
\caption{Statistics of the full dataset across five programming languages.}
\label{tab:full_dataset_stats}
\begin{tabular}{lccccc}
\toprule
\textbf{Language} & \textbf{\# Methods} & \textbf{\# Instances} & \textbf{Train} & \textbf{Valid} & \textbf{Test} \\
\midrule
Java & 419,404 & 667,860 & 534,288 & 66,786 & 66,786 \\
Python & 197,394 & 417,273 & 333,818 & 41,727 & 41,728 \\
JavaScript & 20,560 & 34,539 & 27,631 & 3,453 & 3,455 \\
TypeScript & 36,530 & 64,613 & 51,690 & 6,461 & 6,462 \\
C\# & 60,708 & 119,412 & 95,529 & 11,941 & 11,942 \\
\bottomrule
\end{tabular}
\end{table}

%% file: figures/experimental_setup/leave-one-out_figure.tex
\begin{figure*}[t]
  \centering
  \includegraphics[width=.9\textwidth]{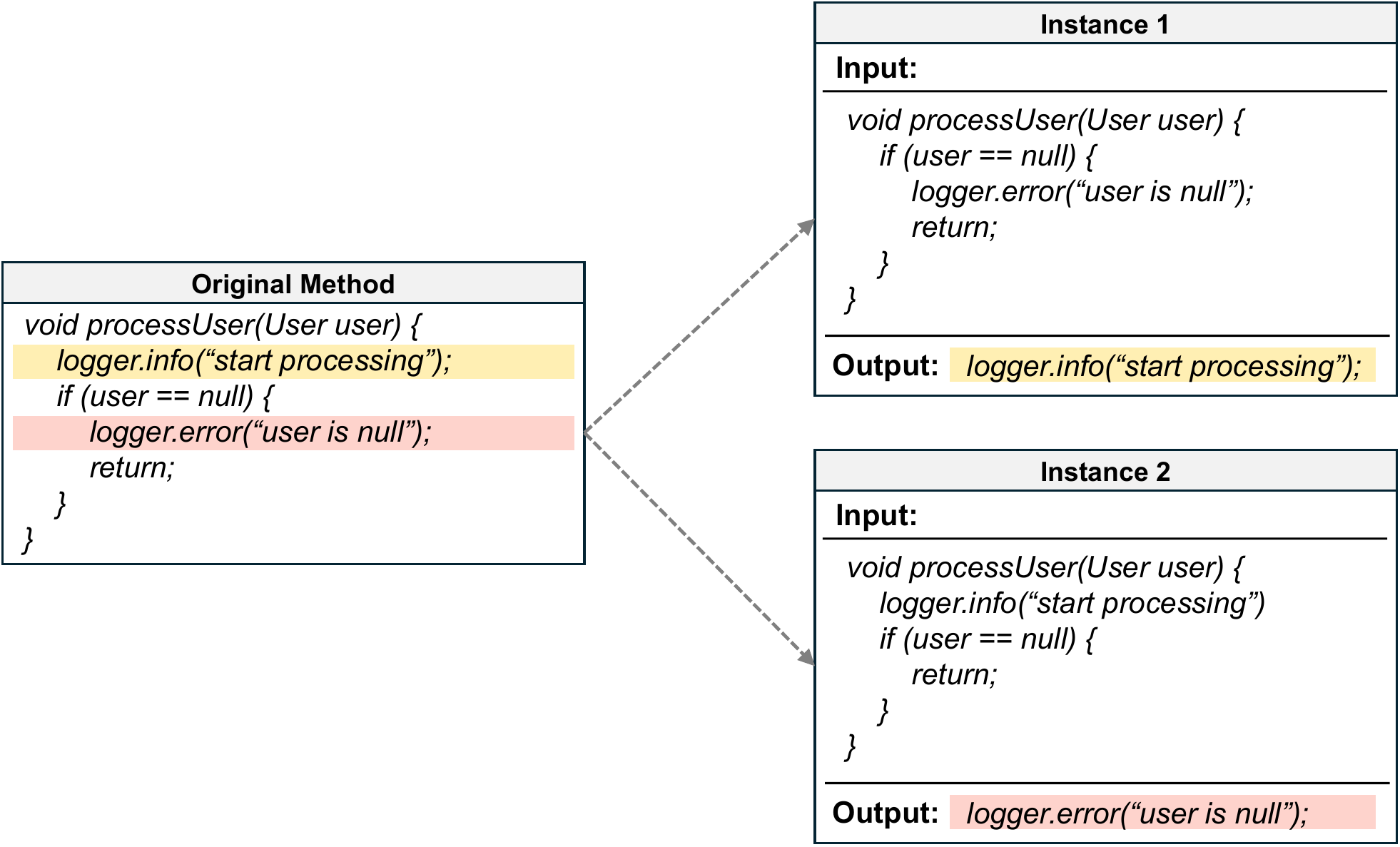}
  \caption{ Illustration of the leave-one-out instance construction process for log statements.}
  \label{fig:leave-one-out}
\end{figure*}

%% file: table/experimentalsetup/dataset_statistics.tex

\newcommand{\levelbar}[1]{%
  \begin{tikzpicture}[baseline=-0.6ex]
    \fill[gray!25] (0,0) rectangle (1.8,0.20);          
    \fill[black]   (0,0) rectangle ({1.8*#1},0.20);     
    \draw[gray!60] (0,0) rectangle (1.8,0.20);          
  \end{tikzpicture}%
}

\begin{table}[t]
\centering
\caption{Log level distribution across five languages. Each cell reports the number of instances (\#), percentage (\%), and an in-cell bar proportional to the percentage.}
\label{tab:dataset_level}
\small
\setlength{\tabcolsep}{4pt}
\resizebox{\textwidth}{!}{%
\begin{tabular}{l|rrc|rrc|rrc|rrc|rrc}
\toprule
\multirow{2}{*}{Level}
& \multicolumn{3}{c|}{Java}
& \multicolumn{3}{c|}{Python}
& \multicolumn{3}{c|}{C\#}
& \multicolumn{3}{c|}{JavaScript}
& \multicolumn{3}{c}{TypeScript}
\\
\cmidrule(lr){2-4}\cmidrule(lr){5-7}\cmidrule(lr){8-10}\cmidrule(lr){11-13}\cmidrule(lr){14-16}
& \# & \% &  & \# & \% &  & \# & \% &  & \# & \% &  & \# & \% &  \\
\midrule
\textit{Critical/Fatal}
& 129   & 0.43 & \levelbar{0.0043}
& 286   & 1.0 & \levelbar{0.0095}
& 339   & 1.1 & \levelbar{0.0113}
& 67    & 0.2 & \levelbar{0.0022}
& 56    & 0.2 & \levelbar{0.0019}
\\
\textit{Error}
& 7,740 & 25.8 & \levelbar{0.2580}
& 4,754 & 15.8 & \levelbar{0.1585}
& 10,183 & 33.9 & \levelbar{0.3394}
& 11,277 & 37.6 & \levelbar{0.3759}
& 9,354 & 31.2 & \levelbar{0.3118}
\\
\textit{Warn}
& 3,824 & 12.7 & \levelbar{0.1274}
& 3,887 & 13.0 & \levelbar{0.1296}
& 3,111 & 10.4 & \levelbar{0.1037}
& 2,996 & 10.0 & \levelbar{0.0999}
& 3,555 & 11.9 & \levelbar{0.1185}
\\
\textit{Info}
& 8,829  & 29.4 & \levelbar{0.2943}
& 14,543 & 48.5 & \levelbar{0.4848}
& 5,537  & 18.5 & \levelbar{0.1846}
& 8,106  & 27.0 & \levelbar{0.2702}
& 8,892  & 29.6 & \levelbar{0.2964}
\\
\textit{Debug}
& 8,276 & 27.6 & \levelbar{0.2759}
& 6,530 & 21.8 & \levelbar{0.2177}
& 7,407 & 24.7 & \levelbar{0.2469}
& 7,120 & 23.7 & \levelbar{0.2373}
& 7,454 & 24.8 & \levelbar{0.2485}
\\
\textit{Trace/Verbose}
& 1,202     & 4.0 & \levelbar{0.0401}
& 0     & 0.0 & \levelbar{0.0000}
& 3,423 & 11.4 & \levelbar{0.1141}
& 434   & 1.4 & \levelbar{0.0145}
& 689   & 2.3 & \levelbar{0.0230}
\\
\midrule
Total
& 30,000 &  & 
& 30,000 &  & 
& 30,000 &  & 
& 30,000 &  & 
& 30,000 &  & 
\\
\bottomrule
\end{tabular}%
}
\end{table}

%% file: section/results.tex
\section{Results} \label{sec:results}
This section presents the experimental results for the three RQs.


\input{section/Results/rq1.tex}
\input{section/Results/rq2.tex}

\input{section/Results/rq3.tex}

%% file: section/Results/rq1.tex
\subsection{RQ1: \RQone} \label{subsec:rq1}
\underline{\textbf{\textit{Approach.}}}
In RQ1, we evaluate the performance of log statement generation approaches in multilingual environments.
Specifically, we compare the performance of existing end-to-end log statement generation approaches and the general-purpose LLMs. 

For the existing end-to-end approaches, we first applied each approach to the training set of each programming language for either fine-tuning or warmup.
Upon completion of this tuning phase, we evaluated the adapted approaches on the unseen test set.
For the LLMs, each model was evaluated under the same task setting using few-shot prompting with data from the same programming language as the target of evaluation.


\underline{\textbf{\textit{Results.}}}
\input{table/results/rq1_results.tex}
Table~\ref{tab:rq1_results} presents the evaluation results for each approach, each programming language, and the average across the five languages.
Boldface indicates the best value for each combination of language and evaluation metric.
Underline indicates the best value among the five LLMs.

\textbf{Observation 1: UniLog achieves the best overall performance among all evaluated approaches in multilingual environments.}
On average across the five languages, UniLog achieved the best values in \textit{Position Accuracy} (75.23\%), \textit{Level Accuracy} (71.26\%), \textit{Message Accuracy} (22.74\%), \textit{BLEU} (26.78), and \textit{All Accuracy} (20.35\%).
\textit{All Accuracy} is the strictest metric because it requires the insertion location, level, and message content to all be correct at the same time.
Even at the language level, UniLog achieved the best \textit{All Accuracy} in all five languages, with 11.87\% for Java, 8.30\% for Python, 44.83\% for JavaScript, 15.17\% for TypeScript, and 21.60\% for C\#.
These results suggest that UniLog not only predicts each element correctly in isolation, but also captures the dependencies among where to insert, which level to assign, and what content to include relatively stably even in multilingual settings.

Among the evaluated LLMs, DeepSeek-V3 achieved the best overall performance.
On average across the five languages, DeepSeek-V3 achieved the highest values among LLMs in \textit{Position Accuracy} (50.62\%), \textit{Level Accuracy} (64.39\%), \textit{Message Accuracy} (13.47\%), \textit{BLEU} (18.74), \textit{ROUGE} (34.22), and \textit{All Accuracy} (10.96\%).
This indicates that DeepSeek-V3 is the most effective LLM for multilingual log statement generation among the evaluated models.
At the same time, UniLog achieves substantially higher \textit{All Accuracy} than DeepSeek-V3, and this difference is statistically significant on the combined 15,000 test instances (20.35\% vs. 10.96\%; difference = 9.39 percentage points; McNemar test, $p < .001$; odds ratio = 8.30).
This result highlights the effectiveness of task-specific design for log statement generation, particularly the combination of retrieval-based prompting and warmup, in multilingual environments.

\textbf{Observation 2: The impact of cross-language differences varies by approach, and an approach that performs well in one language may vary in effectiveness across others.}
The results show that the effect of programming language differs across approaches.
For example, FastLog shows performance relatively close to UniLog in Java, with \textit{Position Accuracy} of 74.20\%, \textit{Level Accuracy} of 60.87\%, and \textit{BLEU} of 19.33, where the \textit{BLEU} score even exceeds that of UniLog at 17.25.
In Python, FastLog achieves only 38.90\% in \textit{Position Accuracy} and 44.03\% in \textit{Level Accuracy}, which are substantially lower than 61.20\% and 62.10\% for UniLog.
These results indicate that even if an approach is competitive in one language, its effectiveness may still vary across other languages.
Overall, these findings suggest that judging the effectiveness of log statement generation approaches based only on results from a single language is insufficient.
In particular, in multilingual environments, the robustness of each approach can vary substantially depending on the target language, and therefore cross-language comparison is essential.

\textbf{Observation 3: The difficulty of log statement generation varies substantially across programming languages.}
Across both existing approaches and LLMs, JavaScript tends to show the highest performance, whereas Python tends to show the lowest performance.
For example, UniLog achieved an \textit{All Accuracy} of 44.83\% for JavaScript, but only 8.30\% for Python.
Similarly, DeepSeek-V3 achieved an \textit{All Accuracy} of 20.27\% for JavaScript, whereas its performance on Python was much lower.
This tendency indicates that prediction difficulty differs substantially across programming languages.

Such differences may be influenced by language-specific factors, such as syntax, implementation practices, logging APIs, and coding styles.
They may also be affected by the number of candidate insertion locations and the degree to which log messages are formulaic.
In particular, Python shows low \textit{Message Accuracy} for many approaches, which suggests that the diversity and context dependence of message content may be a bottleneck for overall performance.
These findings indicate that evaluating log statement generation only on a single programming language is insufficient, and that multilingual evaluation is necessary to understand the practical effectiveness and limitations of each approach.
To further clarify the factors behind these cross-language differences, RQ3 provides a detailed analysis of language-specific characteristics of log statements.

\textbf{Observation 4: Self-hosted LLMs can be competitive with API-based LLMs on some subtasks, but their end-to-end performance remains limited.}
Although DeepSeek-V3 achieved the best overall performance among the evaluated LLMs, self-hosted LLMs also showed competitive performance on some individual metrics.
For example, Qwen2.5-Coder achieved an average \textit{Position Accuracy} of 30.97\%, which is close to GPT-4.1 mini at 35.44\%.
For \textit{Level Accuracy}, Llama3 achieved 52.97\%, whereas GPT-4.1 mini achieved 54.29\%, which are nearly the same.
On message-related metrics, Llama3 achieved a \textit{Message Accuracy} of 12.20\%, \textit{BLEU} of 15.96, and \textit{ROUGE} of 28.93, all of which are higher than 6.25\%, 11.10, and 25.35 for GPT-4.1 mini, respectively.

These results suggest that self-hosted LLMs can be viable options for specific subtasks, especially when cost, privacy, or deployment constraints are important.
Their \textit{All Accuracy} remains lower than that of DeepSeek-V3 and much lower than that of UniLog.
When applying LLMs to multilingual log statement generation, it is important to consider not only overall performance but also subtask-specific strengths and practical deployment constraints.

\begin{tcolorbox}
\textbf{RQ1 Summary:}
In the unified performance comparison of log statement generation approaches in multilingual environments, UniLog achieved the best overall performance among all evaluated approaches, with an average \textit{All Accuracy} of 20.35\% across the five languages.
Among the evaluated LLMs, DeepSeek-V3 achieved the highest performance, with an average \textit{All Accuracy} of 10.96\%.
The comparison between UniLog and DeepSeek-V3 highlights the effectiveness of task-specific design, retrieval-based prompting, and warmup for multilingual log statement generation.
We also observed clear differences in prediction difficulty across programming languages: JavaScript consistently showed the highest performance, whereas Python showed the lowest performance.
Self-hosted LLMs can be competitive with API-based LLMs on some individual metrics, but their end-to-end performance remains limited.
Overall, these results indicate that both the choice of approach and the target programming language strongly affect the performance of multilingual log statement generation.
\end{tcolorbox}

%% file: table/results/rq1_results.tex
\begin{table*}[tbp]
\centering
\small
\caption{The performance of log statement generation approaches in multilingual settings.}
\label{tab:rq1_results}
\begin{tabular}{llcccccc}
\toprule
\multirow{2}{*}{\textbf{Language}} &
\multirow{2}{*}{\textbf{Approach}}  &
\multicolumn{1}{c}{\textbf{Position}} &
\multicolumn{1}{c}{\textbf{Level}}    &
\multicolumn{3}{c}{\textbf{Message}}  &
\multicolumn{1}{c}{\textbf{All}}      \\
\cmidrule(lr){3-3} \cmidrule(lr){4-4} \cmidrule(lr){5-7} \cmidrule(lr){8-8}
& & \textbf{Acc} & \textbf{Acc} & \textbf{Acc} & \textbf{BLEU} & \textbf{ROUGE} & \textbf{Acc} \\
\midrule

\multirow{8}{*}{Java}
& LANCE        & 61.77           & 56.43           & 6.13            & 7.48            & 9.05            & 4.83            \\
& FastLog       & 74.20           & 60.87           & 9.27            & \textbf{19.33}  & \textbf{45.48}  & 7.27            \\
& UniLog       & \textbf{75.97}  & \textbf{67.97}  & \textbf{14.33}  & 17.25           & 34.01           & \textbf{11.87}  \\
& Llama3        & 21.27           & 42.83           & 6.93            & 10.07           & 23.70           & 3.47            \\
& Qwen2.5-Coder & 35.37           & 52.03           & 7.13            & 10.60           & 24.91           & 4.40            \\
& Mistral       & 25.97           & 50.10           & 4.60            & 8.87            & 22.56           & 3.03            \\
& GPT-4.1 mini  & 35.27           & 55.70           & 4.53            & 9.49            & 24.84           & 3.60            \\
& DeepSeek-V3   & \underline{44.67} & \underline{60.77} & \underline{8.20} & \underline{13.08} & \underline{29.66} & \underline{6.30} \\
\midrule

\multirow{8}{*}{Python}
& LANCE        & 39.97           & 43.87           & 3.37            & 3.55            & 4.81            & 2.60            \\
& FastLog       & 38.90           & 44.03           & 3.17            & 8.23            & \textbf{32.94}  & 2.03            \\
& UniLog       & \textbf{61.20}  & \textbf{62.10}  & \textbf{11.33}  & \textbf{13.23}  & 26.76           & \textbf{8.30}   \\
& Llama3        & 9.40            & 43.77           & 4.23            & 4.71            & 12.01           & 1.97            \\
& Qwen2.5-Coder & 19.23           & 43.23           & 4.27            & 5.71            & 15.53           & 1.80            \\
& Mistral       & 13.27           & 49.67           & 2.57            & 5.02            & 13.94           & 0.73            \\
& GPT-4.1 mini  & 20.70           & 51.23           & 1.87            & 4.69            & 14.46           & 0.97            \\
& DeepSeek-V3   & \underline{33.30} & \underline{55.23} & \underline{5.00} & \underline{7.71} & \underline{18.70} & \underline{3.10} \\
\midrule

\multirow{8}{*}{JavaScript}
& LANCE        & 67.67           & 62.07           & 10.77           & 11.83           & 14.57           & 9.73            \\
& FastLog       & 77.70           & 74.40           & 23.17           & 38.41           & 59.98           & 20.90           \\
& UniLog       & \textbf{84.17}  & \textbf{81.93}  & \textbf{46.83}  & \textbf{51.59}  & \textbf{67.08}  & \textbf{44.83}  \\
& Llama3        & 39.00           & 69.97           & \underline{24.40} & 31.49           & 48.31           & 11.83           \\
& Qwen2.5-Coder & 41.40           & 51.13           & 8.07            & 21.81           & 37.84           & 8.07            \\
& Mistral       & 34.60           & 59.83           & 9.80            & 17.30           & 34.01           & 5.20            \\
& GPT-4.1 mini  & 42.37           & 58.63           & 7.20            & 14.39           & 31.76           & 4.07            \\
& DeepSeek-V3   & \underline{69.77} & \underline{76.83} & 23.90 & \underline{32.57} & \underline{50.87} & \underline{20.27} \\
\midrule

\multirow{8}{*}{TypeScript}
& LANCE        & 58.27           & 56.57           & 4.20            & 4.86            & 6.65            & 3.37            \\
& FastLog       & 68.63           & 62.83           & 6.90            & 20.01           & \textbf{44.27}  & 6.23            \\
& UniLog       & \textbf{78.53}  & \textbf{72.83}  & \textbf{17.03}  & \textbf{24.96}  & 43.19           & \textbf{15.17}  \\
& Llama3        & 27.77           & 55.50           & 9.73            & 15.74           & 29.37           & 6.47            \\
& Qwen2.5-Coder & 30.07           & 47.50           & 8.33            & 14.21           & 29.43           & 4.67            \\
& Mistral       & 27.60           & 51.47           & 6.77            & 12.88           & 27.20           & 3.80            \\
& GPT-4.1 mini  & 32.30           & 52.33           & 6.60            & 12.94           & 27.73           & 4.50            \\
& DeepSeek-V3   & \underline{45.47} & \underline{63.17} & \underline{12.03} & \underline{19.23} & \underline{35.19} & \underline{9.20} \\
\midrule

\multirow{8}{*}{C\#}
& LANCE        & 59.03           & 50.50           & 8.93            & 11.18           & 13.41           & 7.60            \\
& FastLog       & 65.37           & 63.93           & 11.67           & 23.02           & \textbf{48.90}  & 10.13           \\
& UniLog       & \textbf{76.27}  & \textbf{71.47}  & \textbf{24.17}  & \textbf{26.85}  & 44.67           & \textbf{21.60}  \\
& Llama3        & 17.40           & 52.80           & 15.70           & 17.80           & 31.27           & 6.33            \\
& Qwen2.5-Coder & 28.77           & 46.87           & 12.80           & 15.45           & 29.50           & 7.13            \\
& Mistral       & 14.00           & 49.30           & 9.33            & 11.98           & 25.71           & 3.67            \\
& GPT-4.1 mini  & 46.57           & 53.57           & 11.07           & 13.99           & 27.95           & 9.13            \\
& DeepSeek-V3   & \underline{59.87} & \underline{65.93} & \underline{18.20} & \underline{21.09} & \underline{36.69} & \underline{15.93} \\
\midrule

\multirow{8}{*}{Average}
& LANCE        & 57.34           & 53.89           & 6.68            & 7.78            & 9.70            & 5.63            \\
& FastLog       & 64.96           & 61.21           & 10.84           & 21.80           & \textbf{46.31}  & 9.31            \\
& UniLog       & \textbf{75.23}  & \textbf{71.26}  & \textbf{22.74}  & \textbf{26.78}  & 43.14           & \textbf{20.35}  \\
& Llama3        & 22.97           & 52.97           & 12.20           & 15.96           & 28.93           & 6.01            \\
& Qwen2.5-Coder & 30.97           & 48.15           & 8.12            & 13.56           & 27.44           & 5.21            \\
& Mistral       & 23.09           & 52.07           & 6.61            & 11.21           & 24.68           & 3.29            \\
& GPT-4.1 mini  & 35.44           & 54.29           & 6.25            & 11.10           & 25.35           & 4.45            \\
& DeepSeek-V3   & \underline{50.62} & \underline{64.39} & \underline{13.47} & \underline{18.74} & \underline{34.22} & \underline{10.96} \\
\bottomrule
\end{tabular}
\end{table*}

%% file: section/Results/rq2.tex
\subsection{RQ2: \RQtwo} \label{subsec:rq2}
\underline{\textbf{\textit{Approach.}}}
To compare the two training strategies (i) Monolingual LoRA (Mono-LoRA) and (ii) Multilingual LoRA (Multi-LoRA), we adopt Llama3 as the backbone model because it achieved the highest average \textit{All Accuracy} among the \textit{self-hosted} models in RQ1. Self-hosted is a necessary condition for adapting the training strategy by LoRA.

For Mono-LoRA, we train a separate LoRA for each language by using 24,000 training instances from each language, and we evaluate the effect of language-specific adaptation.
For Multi-LoRA, we prepare two settings in order to also examine the effect of training data size in multilingual joint training.
The first setting trains a single LoRA on a total of 24,000 instances, with 4,800 instances from each language.
This setting enables a fair comparison of the learning unit itself, monolingual versus multilingual joint training, by matching the total amount of training data with Mono-LoRA.
The second setting trains a single LoRA on a total of 120,000 instances, with 24,000 instances from each language.
This setting evaluates the effect of scaling up the training data in multilingual joint training.

In addition, we also evaluate UniLog and FastLog for comparison with existing approaches.
In particular, although UniLog uses GPT-4.1 mini by default, in this study we evaluate an implementation that uses Llama3 in order to make the comparison with the LoRA-trained models as fair as possible.
We refer to this implementation as UniLog-L.
For UniLog-L, we also construct Mono-UniLog-L, which performs warmup independently for each language, and Multi-UniLog-L, which performs warmup on the integrated data of the five languages, in correspondence with Mono-LoRA and Multi-LoRA, and compare the effect of the learning unit.
Similarly, for FastLog, we construct Mono-FastLog and Multi-FastLog.
Table~\ref{tab:rq2_method_summary} shows the training settings of these approaches.
The Training Setting column indicates whether training is performed monolingually or multilingually.
The Total column indicates the total number of training instances across all five languages, whereas the /Lang. column indicates the number of training instances allocated to each language.
\input{table/results/training_configurations.tex}

\underline{\textbf{\textit{Results.}}}
\input{table/results/rq2_results.tex}
Table~\ref{tab:rq2_results} presents the language-wise performance and the average performance across the five languages for all approaches evaluated in RQ2.
Boldface indicates the best value for each combination of programming language and evaluation metric.
Underline indicates the second-best value.

\textbf{Observation 5: Even a small amount of language-specific warmup can outperform much larger training setups.}
As shown in Table~\ref{tab:rq2_results}, Mono-UniLog-L achieves an average \textit{All Accuracy} of 16.24\% across the five languages, which is the highest among all evaluated approaches.
It also records the best values for \textit{Level Accuracy} at 69.23\%, \textit{Message Accuracy} at 20.29\%, and \textit{BLEU} at 24.81, which indicates that it is particularly strong at predicting log levels and message content.
At the language level, it also achieves the best \textit{All Accuracy}, with 34.77\% for JavaScript and 19.60\% for C\#.
Although Mono-UniLog-L achieves the highest performance in this way, what is noteworthy is its amount of training data.
As shown in Table~\ref{tab:rq2_method_summary}, Mono-UniLog-L is trained with only 500 instances, yet it outperforms Mono-LoRA at 15.13\%, which is trained with 24,000 instances per language, and Multi-LoRA-120k at 15.94\%, which is trained with 120,000 instances in total.
These results indicate that, in multilingual log statement generation, UniLog-style adaptation with a small amount of warmup for each language can achieve both very high data efficiency and strong overall performance.

\textbf{Observation 6: When the total amount of training data is controlled, monolingual training consistently outperforms multilingual training.}
As shown in Table~\ref{tab:rq2_results} and Table~\ref{tab:rq2_method_summary}, when approaches are compared under the same total amount of training data, monolingual training consistently outperforms multilingual training.
For the UniLog-based approaches, Mono-UniLog-L achieves an average \textit{All Accuracy} of 16.24\% across the five languages, which is higher than Multi-UniLog-L at 12.80\%.
Similarly, for the FastLog-based approaches, Mono-FastLog achieves an \textit{All Accuracy} of 9.31\%, which is higher than Multi-FastLog at 5.22\%.
The same tendency is observed for the LoRA-based approaches.
When the total amount of training data is fixed at 24,000 instances, Mono-LoRA achieves an \textit{All Accuracy} of 15.13\%, which is higher than Multi-LoRA-24k at 13.10\%.
This tendency is also consistent with the difference in the number of training instances allocated to each language.
For example, in the LoRA setting, Mono-LoRA is trained with 24,000 instances for each language, whereas Multi-LoRA-24k shares a total of 24,000 instances across the five languages, which means that only 4,800 instances are used for each language.
The same holds for FastLog, where Mono-FastLog uses 24,000 instances per language, whereas Multi-FastLog uses 4,800 instances per language.
For the UniLog-based approaches as well, Mono-UniLog-L uses 500 instances per language, whereas Multi-UniLog-L uses 500 instances in total, which corresponds to 100 instances per language.
These results suggest that, under a fixed total training budget, the disadvantage caused by the reduction in training data for each language outweighs the benefit of sharing information through multilingual integration.
In multilingual log statement generation, monolingual training is more advantageous when the total amount of training data is limited.
This comparison controls for the total amount of training data rather than the full computational or operational cost, since monolingual training requires separate training for each language whereas multilingual training uses a single shared model.

\textbf{Observation 7: Position prediction is more sensitive to training data scale than level and message prediction.}
As shown in Table~\ref{tab:rq2_method_summary} and Table~\ref{tab:rq2_results}, \textit{Position Accuracy} tends to improve overall as the amount of training data for each language increases.
For example, when focusing on the multilingual training settings, the average \textit{Position Accuracy} across the five languages is 51.38\% for Multi-UniLog-L, which uses only 100 instances per language.
It increases to 61.26\% for Multi-FastLog, which uses 4,800 instances per language, to 62.41\% for Multi-LoRA-24k, and to 67.00\% for Multi-LoRA-120k, which uses 24,000 instances per language.
A similar tendency is also observed in monolingual training.
Mono-UniLog-L, which uses 500 instances for each language, achieves a \textit{Position Accuracy} of 61.20\%, whereas Mono-FastLog and Mono-LoRA, which use 24,000 instances for each language, achieve 64.96\% and 67.45\%, respectively.
These results show that increasing the amount of training data is consistently associated with improvement in position prediction.

This tendency can also be confirmed by comparisons within the same approach.
For example, in Multi-LoRA, when the number of training instances for each language is increased from 4,800 to 24,000, \textit{Position Accuracy} improves from 62.41\% to 67.00\%, which is an increase of 4.59 points and the largest gain among all evaluation metrics.
Under the same comparison, \textit{Level Accuracy} improves from 65.97\% to 67.99\%, which is an increase of 2.02 points, \textit{Message Accuracy} improves from 17.45\% to 18.99\%, which is an increase of 1.54 points, \textit{BLEU} improves from 22.08 to 23.42, which is an increase of 1.34, and \textit{ROUGE} improves from 37.36 to 38.78, which is an increase of 1.42.
\textit{All Accuracy} also improves from 13.10\% to 15.94\%, but this improvement is likely supported mainly by the gain in position prediction.

When looking across all approaches, the performance of level and message prediction is not explained only by the amount of training data, and the effect of approach design is also large.
For example, Mono-UniLog-L, which is trained with only 500 instances per language, achieves \textit{Level Accuracy} of 69.23\% and \textit{Message Accuracy} of 20.29\%, which are higher than those of Mono-FastLog and Mono-LoRA, both of which use 24,000 instances per language.
In other words, for log level and message content prediction, increasing the amount of training data has a certain effect, but its influence is not as dominant as it is for position prediction, and the results are more strongly affected by the adaptation strategy and training design of each approach.
Overall, these findings indicate that increasing the amount of training data has a certain effect on all three elements in multilingual log statement generation, but the effect is strongest for deciding where to insert a log statement.
For log level and message content prediction, the design of the approach itself appears to play a more important role in addition to training scale.

\begin{tcolorbox}
\textbf{RQ2 Summary:}
In the comparison of training strategies for multilingual log statement generation, Mono-UniLog-L achieved the best overall performance, with an average \textit{All Accuracy} of 16.24\% across the five languages.
In particular, although Mono-UniLog-L uses only 500 instances for additional training, it achieved higher performance than models using LoRA with 120k instances.
In addition, when compared under the same total training budget, monolingual training consistently outperformed multilingual joint training.
Increasing the amount of training data had a particularly strong effect on \textit{Position Accuracy}.
For example, in Multi-LoRA, increasing the number of instances per language from 4,800 to 24,000 substantially improved \textit{Position Accuracy} from 62.41\% to 67.00\%, whereas the improvements in \textit{Level Accuracy} and \textit{Message Accuracy} were relatively small.
These results indicate that, in multilingual log statement generation, language-specific adaptation is the most effective strategy under a limited training budget.
They also indicate that scaling up training data especially contributes to improving log insertion location prediction, whereas the prediction of log level and message content is more strongly affected by the design of the approach itself.
\end{tcolorbox}

%% file: table/results/training_configurations.tex
\begin{table}[t]
\centering
\caption{Summary of training configurations for the approaches used in RQ2.}
\label{tab:rq2_method_summary}
\begin{tabular}{lccc}
\toprule
\textbf{Approach}  & \textbf{Training Setting} & \textbf{Total} & \textbf{/Lang.} \\
\midrule
Llama3 (no LoRA)   & --          & 0        & 0       \\
Mono-UniLog-L      & Monolingual       & 500      & 500     \\
Multi-UniLog-L     & Multilingual      & 500      & 100     \\
Mono-FastLog       & Monolingual       & 24,000   & 24,000  \\
Multi-FastLog      & Multilingual      & 24,000   & 4,800   \\
Mono-LoRA          & Monolingual       & 24,000   & 24,000  \\
Multi-LoRA-24k     & Multilingual      & 24,000   & 4,800   \\
Multi-LoRA-120k    & Multilingual      & 120,000  & 24,000  \\
\bottomrule
\end{tabular}
\end{table}

%% file: table/results/rq2_results.tex
\begin{table}[tbp]
\centering
\small
\caption{The performance of different training strategies in multilingual log statement generation.}
\label{tab:rq2_results}
\begin{tabular}{llcccccc}
\toprule
\multirow{2}{*}{\textbf{Language}} &
\multirow{2}{*}{\textbf{Methods}}  &
\multicolumn{1}{c}{\textbf{Position}} &
\multicolumn{1}{c}{\textbf{Level}}    &
\multicolumn{3}{c}{\textbf{Message}}  &
\multicolumn{1}{c}{\textbf{All}}      \\
\cmidrule(lr){3-3} \cmidrule(lr){4-4} \cmidrule(lr){5-7} \cmidrule(lr){8-8}
& & \textbf{Acc} & \textbf{Acc} & \textbf{Acc} & \textbf{BLEU} & \textbf{ROUGE} & \textbf{Acc} \\
\midrule

\multirow{8}{*}{Java}
& Llama3 (no LoRA)     & 21.27                & 42.83                & 6.93                  & 10.07                & 23.70                & 3.47                 \\
& Mono-UniLog-L        & 66.90                & \textbf{66.60}       & 11.60                 & \underline{16.42}    & 34.37                & 8.53                 \\
& Multi-UniLog-L       & 59.90                & 63.27                & \textbf{12.53}        & 16.10                & 34.67                & 8.10                 \\
& Mono-FastLog         & \textbf{74.20}       & 60.87                & 9.27                  & \textbf{19.33}       & \textbf{45.48}       & 7.27                 \\
& Multi-FastLog        & 71.20                & 59.83                & 6.47                  & 15.91                & \underline{40.63}    & 4.87                 \\
& Mono-LoRA            & \underline{72.13}    & \underline{65.37}    & \underline{12.37}     & 16.29                & 34.26                & \textbf{10.73}       \\
& Multi-LoRA-24k       & 65.23                & 62.90                & 11.07                 & 15.27                & 32.56                & 8.57                 \\
& Multi-LoRA-120k      & 65.87                & 63.23                & 11.67                 & 15.78                & 33.17                & \underline{9.70}     \\
\midrule

\multirow{8}{*}{Python}
& Llama3 (no LoRA)     &  9.40                & 43.77                &  4.23                 &  4.71                & 12.01                &  1.97                \\
& Mono-UniLog-L        & 35.27                & 56.50                & \underline{9.17}      & 10.77                & 22.64                &  5.07                \\
& Multi-UniLog-L       & 30.90                & 56.20                &  8.33                 &  9.91                & 21.31                &  4.43                \\
& Mono-FastLog         & 38.90                & 44.03                &  3.17                 &  8.23                & \textbf{32.94}       &  2.03                \\
& Multi-FastLog        & 37.33                & 47.53                &  2.87                 &  7.92                & 30.11                &  1.70                \\
& Mono-LoRA            & \textbf{52.93}       & \textbf{57.73}       & \textbf{9.93}         & \underline{11.44}    & \underline{22.83}    & \underline{6.90}     \\
& Multi-LoRA-24k       & 50.17                & 56.37                &  9.03                 & 10.81                & 21.67                &  6.27                \\
& Multi-LoRA-120k      & \underline{52.30}    & \underline{57.67}    & \textbf{9.93}         & \textbf{11.46}       & 22.38                & \textbf{7.23}        \\
\midrule

\multirow{8}{*}{JavaScript}
& Llama3 (no LoRA)     & 39.00                & 69.97                & 24.40                 & 31.49                & 48.31                & 11.83                \\
& Mono-UniLog-L        & 73.30                & \textbf{82.40}       & \textbf{41.70}        & \textbf{48.04}       & \textbf{64.56}       & \textbf{34.77}       \\
& Multi-UniLog-L       & 68.53                & \underline{79.37}    & \underline{34.97}     & \underline{41.55}    & 57.89                & 27.50                \\
& Mono-FastLog         & \textbf{77.70}       & 74.40                & 23.17                 & 38.41                & 59.98                & 20.90                \\
& Multi-FastLog        & 70.80                & 64.13                & 12.00                 & 22.96                & 45.74                & 10.43                \\
& Mono-LoRA            & \underline{75.20}    & 78.30                & 33.47                 & 40.74                & 58.26                & \underline{28.30}    \\
& Multi-LoRA-24k       & 63.10                & 76.53                & 32.63                 & 40.55                & 57.95                & 22.50                \\
& Multi-LoRA-120k      & 73.27                & 79.77                & 36.03                 & 43.27                & \underline{60.37}    & 30.07                \\
\midrule

\multirow{8}{*}{TypeScript}
& Llama3 (no LoRA)     & 27.77                & 55.50                &  9.73                 & 15.74                & 29.37                &  6.47                \\
& Mono-UniLog-L        & 64.63                & \textbf{71.90}       & \textbf{15.43}        & \textbf{22.72}       & \underline{38.62}    & \underline{13.23}    \\
& Multi-UniLog-L       & 53.83                & 68.30                & 13.03                 & 20.14                & 35.50                &  9.87                \\
& Mono-FastLog         & 68.63                & 62.83                &  6.90                 & 20.01                & \textbf{44.27}       &  6.23                \\
& Multi-FastLog        & 64.20                & 58.13                &  3.60                 & 14.50                & 38.42                &  3.13                \\
& Mono-LoRA            & \textbf{73.77}       & 69.63                & \underline{14.47}     & \underline{21.71}    & 38.13                & 13.03                \\
& Multi-LoRA-24k       & 68.57                & 68.13                & 13.73                 & 20.50                & 36.65                & 11.93                \\
& Multi-LoRA-120k      & \underline{72.97}    & \underline{70.07}    & 14.80                 & 21.76                & 37.78                & \textbf{13.73}       \\
\midrule

\multirow{8}{*}{C\#}
& Llama3 (no LoRA)     & 17.40                & 52.80                & 15.70                 & 17.80                & 31.27                &  6.33                \\
& Mono-UniLog-L        & \underline{65.90}    & \underline{68.73}    & \textbf{23.57}        & \textbf{26.11}       & \underline{42.05}    & \textbf{19.60}       \\
& Multi-UniLog-L       & 43.73                & 66.63                & \underline{20.97}     & \underline{23.35}    & 38.86                & 14.10                \\
& Mono-FastLog         & 65.37                & 63.93                & 11.67                 & 23.02                & \textbf{48.90}       & 10.13                \\
& Multi-FastLog        & 62.77                & 56.70                &  7.27                 & 15.43                & 41.31                &  5.97                \\
& Mono-LoRA            & 63.23                & 66.77                & 20.47                 & 23.18                & 38.06                & \underline{16.67}    \\
& Multi-LoRA-24k       & 65.00                & 65.93                & 20.77                 & 23.26                & 37.97                & 16.23                \\
& Multi-LoRA-120k      & \textbf{70.60}       & \textbf{69.23}       & \underline{22.53}     & \underline{24.81}    & 40.19                & \underline{18.97}    \\
\midrule

\multirow{8}{*}{Average}
& Llama3 (no LoRA)     & 22.97                & 52.97                & 12.20                 & 15.96                & 28.93                & 6.01                 \\
& Mono-UniLog-L        & 61.20                & \textbf{69.23}       & \textbf{20.29}        & \textbf{24.81}       & \underline{40.45}    & \textbf{16.24}       \\
& Multi-UniLog-L       & 51.38                & 66.75                & 17.97                 & 22.21                & 37.65                & 12.80                \\
& Mono-FastLog         & 64.96                & 61.21                & 10.84                 & 21.80                & \textbf{46.31}       & 9.31                 \\
& Multi-FastLog        & 61.26                & 57.26                & 6.44                  & 15.34                & 39.24                & 5.22                 \\
& Mono-LoRA            & \textbf{67.45}       & 67.56                & 18.14                 & 22.67                & 38.31                & 15.13                \\
& Multi-LoRA-24k       & 62.41                & 65.97                & 17.45                 & 22.08                & 37.36                & 13.10                \\
& Multi-LoRA-120k      & \underline{67.00}    & \underline{67.99}    & \underline{18.99}     & \underline{23.42}    & 38.78                & \underline{15.94}    \\
\bottomrule

\end{tabular}
\end{table}

%% file: section/Results/rq3.tex
\subsection{RQ3: \RQthree} \label{subsec:rq3}

\underline{\textbf{\textit{Approach.}}}
\input{table/results/rq3/category.tex}
In RQ3, we analyze whether the cross-language performance differences observed in RQ1 can be explained by language-specific logging characteristics.
Specifically, we use log insertion categories as the main axis of analysis and examine the three elements that constitute log statement generation: insertion location, log level, and log message.

To conduct a systematic analysis, we classify log insertion locations into six categories and use these categories to analyze position, level, and message characteristics across languages.
The category definitions are based on the framework proposed by Li et al.~\cite{Li2021ASE}, which systematized log insertion locations in Java methods, and we adopt the following six categories: Try-Catch Block, Branching Block, Looping Block, Method Start, Method End, and Domain-Specific Methods.
The definitions of these six categories are shown in Table~\ref{tab:log_position}.
In the following analysis, we analyze the characteristics of log insertion locations, levels, and message content on the basis of these categories.

The category classification was performed with GPT-5.2~\cite{GPT5-2} on all 15{,}000 instances in the test set.
We selected GPT-5.2 because it was the most advanced model available through the OpenAI API at the time of the experiments.
Such state-of-the-art LLMs have shown strong performance in annotating software engineering artifacts~\cite{li2025arXiv,ahmed2025MSR}, making them suitable for this task.
Because the validity of this classification affects the subsequent analysis, we manually verified the classification results.
Assuming a population size of 15{,}000, a 95\% confidence level, and a margin of error of $\pm5\%$, we calculated a required sample size of 375 and randomly sampled these instances.
The first and second authors independently assigned categories to the sampled instances, and the inter-annotator agreement reached a Cohen's kappa of 0.95.
When their labels differed, they resolved the disagreements through discussion and determined the final adjudicated labels.
We then compared the GPT-5.2-assigned labels with these adjudicated labels and found an agreement rate of 88.8\%.
These results indicate that the GPT-5.2-based category classification is sufficiently reliable for the RQ3 analysis.

For the accuracy analysis in RQ3, we use the prediction results of DeepSeek-V3, which is the LLM that achieved the highest performance in RQ1.

\subsubsection{\textbf{Results for Position}}
\input{figures/RQ3/Position/heat_Position_Acc_figure.tex}
In this subsection, we analyze factors that cause the prediction difficulty of log insertion location (i.e., Position) to differ across languages.
Figure~\ref{fig:rq3_pos_heatmap} shows \textit{Position Accuracy} for each combination of programming language, which includes five languages and ALL, and position category.

\textbf{Observation 8: The difficulty differs substantially across position categories, and in particular Category~3, which is Looping Block, is systematically the most difficult.}
When focusing on the \textit{ALL} row in Figure~\ref{fig:rq3_pos_heatmap}, the difficulty of position prediction differs substantially across categories.
The lowest accuracy is observed for Category~3, which is Looping Block, where \textit{Position Accuracy} remains at 26.3\%.
Next, Category~5, which is Method End, is also low at 32.6\%, which indicates that identifying insertion locations is difficult even near the end of a method.
Category~1, which is Try-Catch Block, and Category~6, which is Domain-Specific Methods, are relatively high, at 56.4\% and 57.5\%, respectively.
These results show that, even for the same position prediction task, the difficulty systematically changes depending on structural context.
These findings suggest that position prediction is not uniformly difficult, and that in contexts such as loops and method ends, where multiple reasonable candidate locations are likely to exist, insertion ambiguity increases and the task becomes inherently more difficult for models.

\textbf{Observation 9: The distribution of position categories differs across languages, and the proportion of difficult categories varies across languages.}
\input{figures/RQ3/Position/cat_distribution_lang_heat_figure.tex}
Next, we examine the extent to which cross-language differences in accuracy can be explained by distributional factors.
Specifically, the proportion of difficult categories contained in each language.
Figure~\ref{fig:rq3_pos_catdist} shows the occurrence ratios of Category~1 to Category~6 in each language.
For example, in Python, the proportions of Category~3, which is the most difficult, and Category~5, which is also low accuracy, are both 5.6\%, which are higher than those in ALL, where Category~3 is 2.8\% and Category~5 is 3.0\%.
In JavaScript, Category~3 is 1.2\% and Category~5 is 1.3\%, which indicates relatively lower exposure to difficult categories.
This difference indicates that category distributions can contribute to cross-language differences in the difficulty of position prediction.


\textbf{Observation 10: Loop density is negatively associated with position accuracy and can serve as an auxiliary factor that explains residuals.}
\input{table/results/rq3/loop_density_by_lang.tex}
We further focus on loop density, which is denoted as \textit{loop\_per\_line}, as a lightweight structural factor that can help explain cross-language differences in the difficulty of position prediction.
Loop density is a metric that represents the average number of loop structures per line of code.
Table~\ref{tab:rq3_loop_density_by_lang} shows the average loop density in each language and the proportion of methods that contain loops.
Our analysis found a moderate negative correlation between \textit{loop\_per\_line} and \textit{Position Accuracy}, with Pearson $r=-0.46$.
This is consistent with Observation 8, which indicates that Category~3, which is Looping Block, is the most difficult category.
As shown in Table~\ref{tab:rq3_loop_density_by_lang}, Python has the highest average loop density, at 0.341, and also has the highest proportion of methods that contain loops, at 23.0\%.
JavaScript has the lowest average loop density, at 0.075, and also has a small proportion of methods that contain loops, at 6.30\%.
In this way, the fact that Python, which has large residuals that are not explained by category distributions, also has high loop density suggests that having more loop structures can be one factor that increases the difficulty of position prediction.

\begin{tcolorbox}
\textbf{Implication (Position):}
We found that cross-language differences in position prediction arise from a combination of (i) category-specific difficulty differences, (ii) differences in category distributions across languages, and (iii) language-specific factors, such as the prevalence of loop structures.
In particular, we found that contexts with more loop structures, which correspond to Category~3, increase the difficulty of position prediction, and that Python contains this category relatively more frequently, which can be one factor behind the difficulty of position prediction in Python.
Improving position prediction accuracy requires considering both strengthening structural representations with an emphasis on difficult categories, especially Category~3, and language adaptation that incorporates language-specific syntax and practices.
\end{tcolorbox}

\subsubsection{\textbf{Results for Level}}
\input{figures/RQ3/Level/heat_Level_Acc_figure.tex}

In this subsection, we analyze factors that cause the prediction difficulty of log levels, which are represented in six bands, to differ across languages.
Figure~\ref{fig:rq3_level_heatmap} shows \textit{Level Accuracy} for each combination of language, which includes five languages and ALL, and position category.

\textbf{Observation 11: Cross-language differences in \textit{Level Accuracy} remain, but category dependence is weaker than for Position.}
When focusing on the \textit{ALL} row in Figure~\ref{fig:rq3_level_heatmap}, \textit{Level Accuracy} is generally within the range of 60.5--72.4\%, and there is no extreme difference in difficulty across categories, unlike what was observed for Position.
Cross-language differences are clear.
JavaScript shows high accuracy across all categories, at 63.9--79.7\%, whereas Python is particularly low for Category~1, which is Try-Catch Block, and Category~2, which is Branching Block, at 40.2\% and 46.1\%, respectively.
This tendency suggests that the difficulty of level prediction is not determined sufficiently by category information alone, and may be strongly affected by language-specific factors, such as operational conventions and practices in each language.

\textbf{Observation 12: The distribution of log levels within the same position category differs across languages.}
\input{figures/RQ3/Level/heat_cat_levelband_Python_vs_JavaScript_figure.tex}
Observation 11 confirms that there is a gap in level prediction accuracy between Python and JavaScript for Category~1 and Category~2.
By comparing the level-band distributions in Category~1 and Category~2, we analyze whether one factor behind the cross-language difference is a difference in the typical levels within the same category.
Figure~\ref{fig:rq3_level_cat_levelband_heatmap} shows the level-band distributions for Category~1 and Category~2 in Python and JavaScript.
For Category~1, \textit{Error} is dominant in JavaScript at 78.8\%, whereas in Python it accounts for only 44.7\%, and \textit{Warn} at 25.2\%, \textit{Debug} at 15.4\%, and \textit{Info} at 13.6\% are also mixed in.
For Category~2 as well, \textit{Error} accounts for more than half in JavaScript, at 50.5\%, whereas in Python it is much smaller at 18.3\%, and \textit{Info} at 33.9\% and \textit{Warn} at 32.1\% are the major bands.
In this way, even within the same category, the typical level can differ by language.
As a result, estimation based only on category information is less likely to work well in Python, which may amplify cross-language differences.

\begin{tcolorbox}
\textbf{\textit{Implication (Level):}}
Because the frequent log levels differ across categories, and because the typical level can also differ across languages even within the same category, the difficulty of level prediction may be more strongly affected by language-specific factors than by category dependence.
For example, in categories with clear cross-language differences, such as Category~1, which is Try-Catch Block, and Category~2, which is Branching Block, a uniform language-independent heuristic is likely to be insufficient, and language adaptation that reflects language-specific operational conventions and practices may be effective.
\end{tcolorbox}

\subsubsection{\textbf{Results for Message}}
\input{figures/RQ3/Message/heat_BLEU_A_figure.tex}

In this subsection, we analyze factors that cause the prediction difficulty of log message content (i.e., Message) to differ across languages.
Because exact matching, \textit{Message Accuracy}, does not sufficiently capture semantically similar generations, we use \textit{BLEU}, which is based on surface similarity, as the main metric in this analysis.
Figure~\ref{fig:rq3_msg_bleu_heatmap} shows \textit{BLEU} for each combination of language, which includes five languages and ALL, and position category.

\textbf{Observation 13: Message prediction shows the largest cross-language differences among the three elements, and the same tendency appears for both exact matching and semantic similarity.}
As shown in Figure~\ref{fig:rq3_msg_bleu_heatmap}, \textit{BLEU} differs substantially across languages, and the cross-language gap is larger than that observed for position and level.
JavaScript shows particularly high scores for Category~1 and Category~2, with 41.7 for Category~1 and 37.5 for Category~2, which indicates that it is easier to generate surface expressions that are close to the reference messages even within the same category.
Python is consistently low across all categories.
These results indicate that message prediction strongly depends on language-specific expression habits and lexical choices, and that message prediction is the element in which cross-language differences are most strongly amplified.

\textbf{Observation 14: There are also difficulty differences across categories, and Category~3, which is Looping Block, is the most difficult even for message prediction.}
When focusing on the \textit{ALL} row in Figure~\ref{fig:rq3_msg_bleu_heatmap}, we can observe difficulty differences across categories.
Although Category~1 shows the highest BLEU score, at 20.4, Category~3 remains the lowest, at 9.7.
This is consistent with the result for Position, where Category~3 was also the most difficult category.
One possible reason is that, within loops, the purpose and situation of logging become more diverse, which makes the message content that should be written more situation-dependent.


\textbf{Observation 15: Message diversity shows a negative correlation with accuracy.}
\input{figures/RQ3/Message/msg_distinct2_figure.tex}
To analyze language-specific factors behind cross-language differences in message prediction accuracy, we compute \textit{distinct2} to quantify how template-like the log messages are (i.e., repetitiveness).
\textit{distinct2} is widely used as a diversity metric in natural language generation~\cite{Li2016NAACL,bao2020ACL,chen2022ACL}, and it represents the proportion of unique consecutive 2-grams in the ground-truth messages.
A higher value indicates greater message diversity, which means lower template-likeness, whereas a lower value indicates a stronger tendency toward template-like and repetitive messages.
Figure~\ref{fig:rq3_msg_distinct2_heatmap} shows \textit{distinct2} for each language and position category.
As shown in Figure~\ref{fig:rq3_msg_distinct2_heatmap}, Category~1 and Category~2 in JavaScript, where high BLEU scores were observed, have particularly low \textit{distinct2}, which indicates that the messages tend to be more formulaic.
Python shows high \textit{distinct2} in many categories, which indicates that the log messages tend to be more diverse.
These findings are consistent with the difficulty of message prediction in Python and with the high accuracy observed for Category~1 and Category~2 in JavaScript.

\begin{tcolorbox}
\textbf{\textit{Implication (Message):}}
We found that cross-language differences in message prediction are strongly related to the diversity of log messages, where higher diversity is associated with lower accuracy.
In particular, Python has diverse messages, which suggests that simple template learning or matching frequent phrases is unlikely to achieve sufficient performance.
JavaScript has relatively formulaic messages in some categories, which makes it easier to achieve high accuracy.
For future automated log message generation, it may be effective to adapt the approach to each language and category by (i) leveraging templating and phrase learning for repetitive categories, while (ii) combining abstraction, such as handling identifiers and values, and retrieval of similar examples, such as RAG, for languages with high diversity, especially Python.
\end{tcolorbox}

%% file: table/results/rq3/category.tex
\begin{table}[t]
\centering
\small
\caption{Six categories of logging locations}
\label{tab:log_position}
\begin{tabular}{lll}
\toprule
\textbf{ID} & \textbf{Category} & \textbf{Description} \\
\midrule
Category~1 & Try-Catch Block & Within a \texttt{try-catch} structure, including a \texttt{try} block or a \texttt{catch} block. \\
Category~2 & Branching Block & Within a branching structure, including blocks of \texttt{if}, \texttt{else}, \texttt{switch}, etc. \\
Category~3 & Looping Block & Within an iteration structure, including blocks of \texttt{for}, \texttt{while}, \texttt{do}, etc. \\
Category~4 & Method Start & Near the beginning of a method body. \\
Category~5 & Method End & Near the end of a method body. \\
Category~6 & Domain-Specific Methods & Within a domain-specific method, such as a method for handling a specific request. \\
\bottomrule
\end{tabular}
\end{table}

%% file: figures/RQ3/Position/heat_Position_Acc_figure.tex
\begin{figure}[t]
  \centering
  \includegraphics[width=0.5\textwidth]{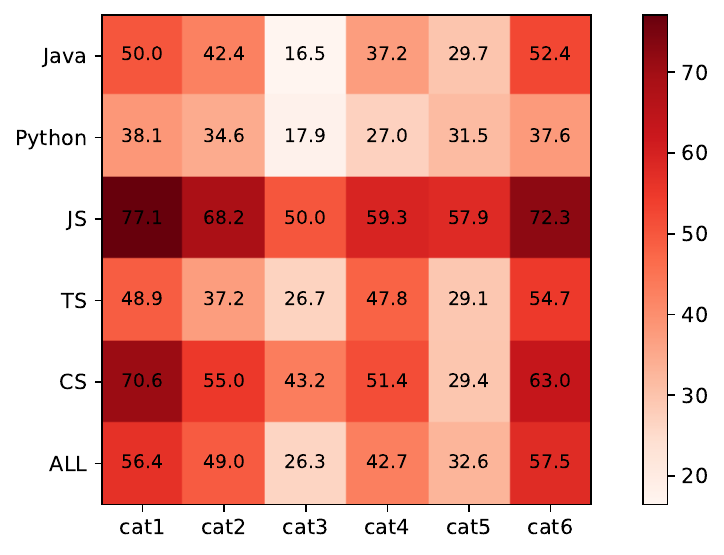}
  \caption{Heatmap of Position Acc. (\%) across Languages and Categories.}
  \label{fig:rq3_pos_heatmap}
\end{figure}

%% file: figures/RQ3/Position/cat_distribution_lang_heat_figure.tex
\begin{figure}[t]
  \centering
  \includegraphics[width=0.5\textwidth]{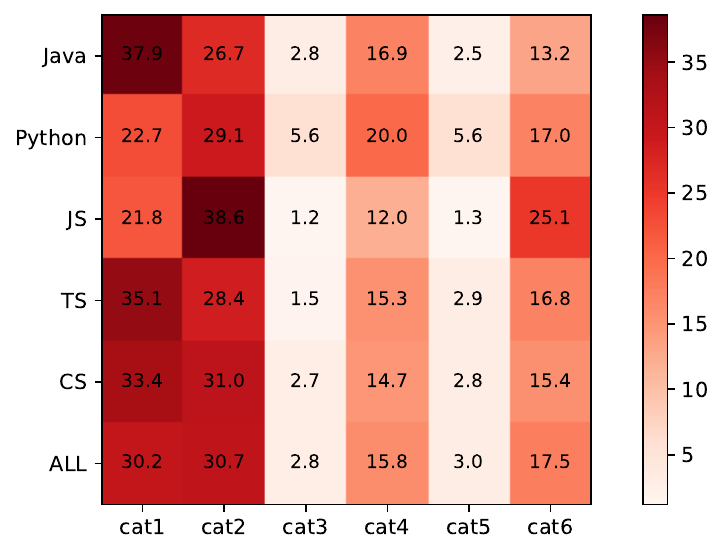}
  \caption{Heatmap of Category Distribution across Languages and Categories.}
  \label{fig:rq3_pos_catdist}
\end{figure}

%% file: table/results/rq3/loop_density_by_lang.tex
\begin{table}[t]
\centering
\caption{Loop density statistics by programming language. Each row reports the average loop density (\texttt{avg\_loop\_per\_line}) and the proportion of methods containing at least one loop (\texttt{loop\_present\_rate}).}
\label{tab:rq3_loop_density_by_lang}
\begin{tabular}{lrr}
\toprule
\textbf{Language} & \textbf{avg\_loop\_per\_line} & \textbf{loop\_present\_rate (\%)} \\
\midrule
Java        & 0.289 & 21.33 \\
Python      & 0.341 & 23.00 \\
JavaScript  & 0.075 &  6.30 \\
TypeScript  & 0.076 &  6.00 \\
C\#         & 0.147 & 11.67 \\
All         & 0.185 & 13.62 \\
\bottomrule
\end{tabular}
\end{table}

%% file: figures/RQ3/Level/heat_Level_Acc_figure.tex
\begin{figure}[t]
  \centering
  \includegraphics[width=0.5\textwidth]{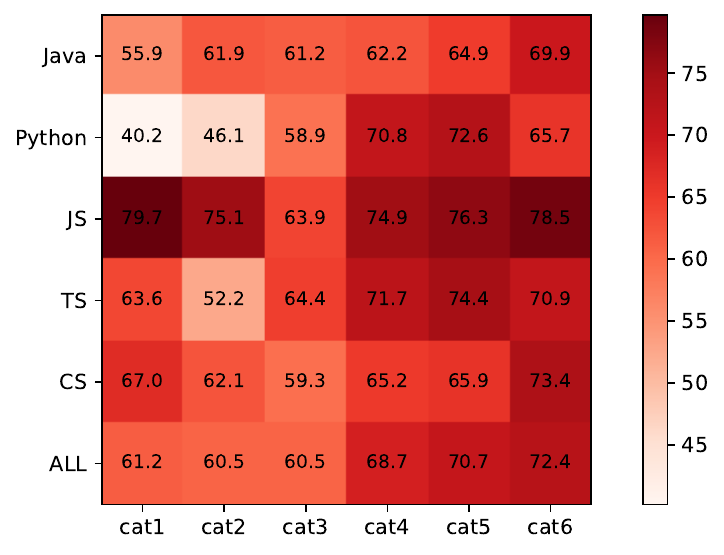}
  \caption{Heatmap of Level Acc. (\%) across Languages and Categories.}
  \label{fig:rq3_level_heatmap}
\end{figure}

%% file: figures/RQ3/Level/heat_cat_levelband_Python_vs_JavaScript_figure.tex
\begin{figure}[t]
  \centering
  \includegraphics[width=\textwidth]{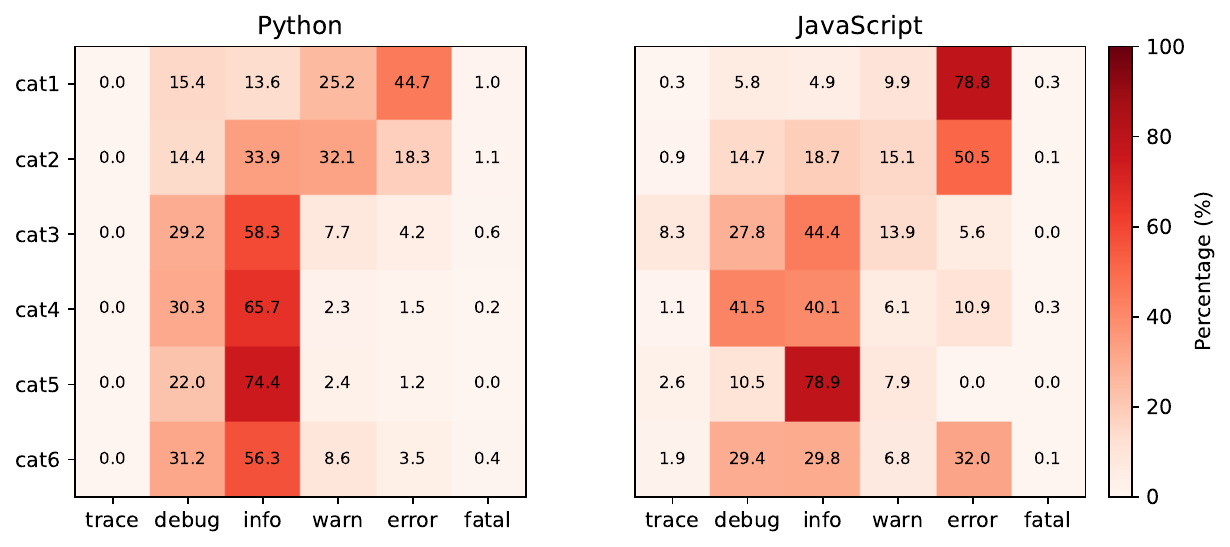}
  \caption{Ground-truth log level-band distributions by position category for Python (left) and JavaScript (right).}
  \label{fig:rq3_level_cat_levelband_heatmap}
\end{figure}

%% file: figures/RQ3/Message/heat_BLEU_A_figure.tex
\begin{figure}[t]
  \centering
  \includegraphics[width=0.5\textwidth]{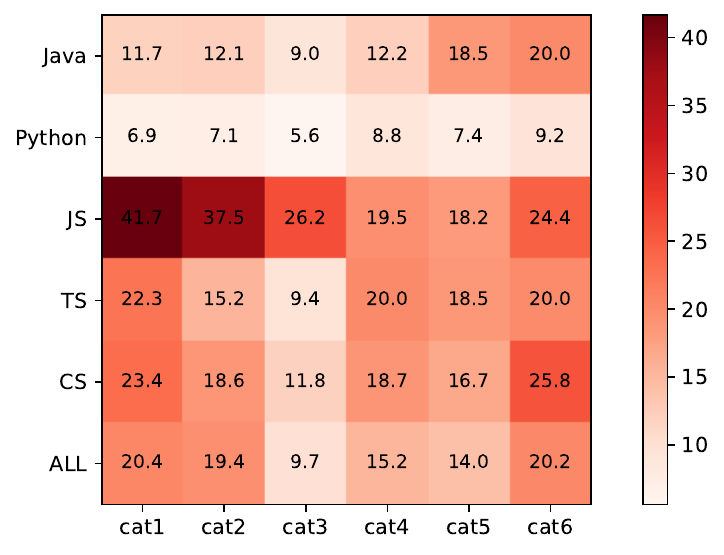}
  \caption{Heatmap of BLEU score across Languages and Categories.}
  \label{fig:rq3_msg_bleu_heatmap}
\end{figure}

%% file: figures/RQ3/Message/msg_distinct2_figure.tex
\begin{figure}[t]
  \centering
  \includegraphics[width=0.5\textwidth]{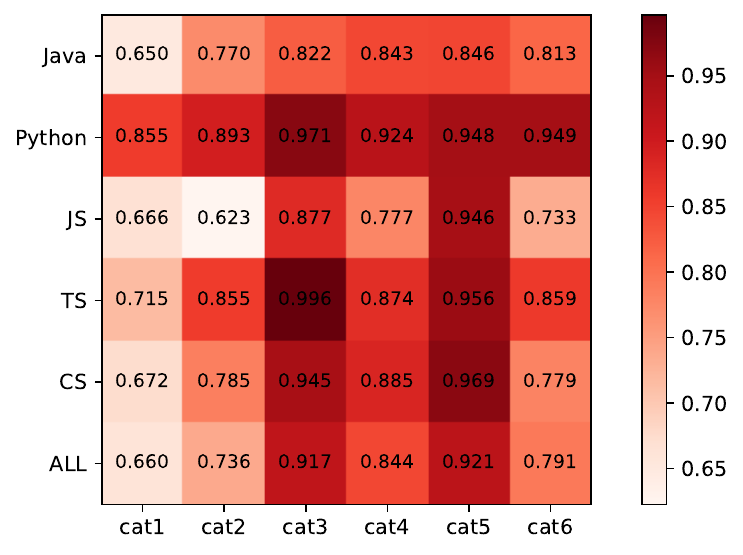}
  \caption{Heatmap of \textit{distinct2} for ground-truth log message diversity across languages and categories.}
  \label{fig:rq3_msg_distinct2_heatmap}
\end{figure}

%% file: section/discussion.tex
\section{Discussion} \label{sec:discussion}
In this section, we discuss three additional analyses that complement the findings of the main RQs.
First, motivated by the results on training strategies, we analyze cross-language generalization in log statement generation to examine how well knowledge learned from one programming language transfers to another.
Second, we investigate the effectiveness of LLMs in a more realistic setting where multiple log statements may appear within a single function, which is not captured by the single-log insertion setting used in the main experiments.
Third, because exact-match and n-gram-based metrics do not fully capture the semantic quality of generated log messages, we conduct a qualitative analysis with LLM-as-a-Judge.

\subsection{Cross-Language Generalization in Log Statement Generation} \label{subsec:cross_language_generalization}
\input{figures/discussion/cross_language_perfect_heatmap_figure.tex}
\input{table/discussion/target_wise.tex}

\underline{\textbf{\textit{Motivation.}}}
In the experiments so far, we have mainly evaluated the effectiveness of LoRA models under language-specific training settings for each programming language.
In real development environments, securing sufficient training data for every language is often difficult.
Clarifying the extent to which a model trained on one language can generalize to other languages is important for considering the practicality of LLM-based log statement generation approaches.
In this section, we analyze cross-language generalization in log statement generation to investigate how beneficial training on the target language is and whether knowledge acquired from one language can be transferred to another language.

\underline{\textbf{\textit{Approach.}}}
To investigate cross-language generalization, we use the five Mono-LoRA models introduced in RQ2.
These models are LoRA models fine-tuned separately for each programming language.
Each LoRA adapter retains adaptation knowledge obtained from a single training language, making these models suitable for analyzing the extent to which such knowledge transfers to other languages.
Llama3 is used as the backbone model for these models.
Each Mono-LoRA model is obtained by independently fine-tuning a LoRA adapter using 24,000 training instances from one programming language.

In this analysis, we evaluate these five Mono-LoRA models on the test sets of all five languages.
We regard the language used for LoRA fine-tuning as the training language and the language used for evaluation as the test language.
This design yields 25 combinations of training and test languages across Java, Python, JavaScript, TypeScript, and C\#, enabling us to systematically examine how differences in the training language affect performance on each target language.
For each combination, we evaluate the generated log statements using All Accuracy.

We distinguish between the in-language setting, where the training and test languages are identical, and the cross-language setting, where they are different.
This comparison quantifies the benefit of training directly on the target language compared with transferring from other languages.
For each target language, we compare the in-language performance with the average performance of the cross-language models trained on the other four languages.
For each target language, we also identify the Best Source: the source language that achieves the highest performance among the cross-language models.
This allows us to analyze not only whether cross-language transfer is possible, but also which training source language transfers most effectively to each target language.

\underline{\textbf{\textit{Results.}}}
Figure~\ref{fig:discussion_cross_language_perfect} shows a heatmap of All Accuracy for each combination of training language and test language.
Table~\ref{tab:cross_language_target_summary} compares, for each target language, the in-language performance with the cross-language average.
Here, in-language refers to the performance of the model trained on the same language as the target language, and cross-language average refers to the average performance of models trained on the other four languages.
Diff represents the difference obtained by subtracting the cross-language average from the in-language performance, and Best Source indicates the source language that achieved the highest performance in the cross-language setting.

\textbf{In-language fine-tuning achieves the highest All Accuracy for most target languages.}
For four languages except C\#, the in-language setting, where the training language and test language are identical, achieved the highest All Accuracy, as shown in Figure~\ref{fig:discussion_cross_language_perfect}.
This result suggests that cross-language transfer is effective to some extent, while models trained directly on the target language are basically the most effective for log statement generation.

\textbf{The advantage of in-language fine-tuning differs substantially across target languages.}
The absolute level of performance differs depending on the target language.
For example, JavaScript shows relatively high values overall, whereas Python remains at relatively low values overall.
A target-wise comparison is therefore needed to evaluate the advantage of the in-language setting fairly.
As shown in Table~\ref{tab:cross_language_target_summary}, the in-language model outperformed the average of the cross-language models for all target languages.
The most notable case was JavaScript, where the in-language model achieved an All Accuracy of 28.30 and outperformed the cross-language average of 22.19 by 6.11 points.
This result suggests that, in JavaScript, language-specific knowledge and logging practices obtained from training data in the same language are particularly important.
For Java, Python, and TypeScript, the differences were only 1.89, 1.42, and 1.75 points, respectively, indicating that the advantage of in-language training was limited.
For C\#, the in-language performance was 16.67, whereas the cross-language average was 16.65, and the difference was only 0.02 points.
This indicates that, for C\#, models trained on other languages can achieve almost equivalent performance.

\textbf{Cross-language transferability is not explained solely by simple language similarity.}
From the perspective of Best Source, cross-language transfer may be influenced to some extent by closeness between languages.
For example, the best cross-language source for JavaScript was TypeScript, which is likely to reflect the fact that JavaScript and TypeScript are close in terms of syntax and development style.
We also observed that Java was the Best Source for Python and Python was the Best Source for Java, which indicates that generalization performance is not explained only by simple language similarity.
Transferability in log statement generation is likely affected not only by superficial syntactic similarity but also by structural characteristics of log insertion locations, such as branching, exception handling, and state updates, as well as by biases in the log patterns contained in each language dataset.

These results indicate that in-language fine-tuning is overall the most effective strategy for log statement generation.
The magnitude of its advantage differs substantially across target languages.
In JavaScript, the effect of in-language adaptation was clearly observed.
In C\#, sufficiently competitive performance was obtained even with cross-language transfer alone.
This suggests that large-scale language-specific data may be unnecessary in some cases.
The results also show that, for some languages, adaptation specialized to the target language remains important.
In the future, in addition to training separate models for each language, an important challenge will be how to balance logging knowledge that can be shared across languages with knowledge that is specific to each target language.

\subsection{Effectiveness of LLMs in Multi-Log-Statement Settings} \label{subsec:multi_log_statement_settings}
\input{table/discussion/multi-dataset.tex}
\input{table/discussion/multi_log_results.tex}

\underline{\textbf{\textit{Motivation.}}}
In the experiments so far, we have evaluated the performance of LLMs for log statement generation under the setting where one log statement is inserted into one function.
In actual development environments, it is common for multiple log statements to be inserted within a single function.
Evaluation that targets only single-log insertion may fail to capture the practical difficulty of log statement generation.
In a setting where a function may contain multiple log statements, the model needs to determine not only where to insert log statements, but also how many log statements should be inserted and how their roles should be differentiated within the same function.
In this section, we additionally analyze a setting in which a single function may contain multiple log statements, as a more realistic setting.

\underline{\textbf{\textit{Approach.}}}
To analyze this setting, we constructed a new dataset by following the same procedure as in Section~\ref{sec:benchmarkconstruction}.
For each language, we collected 200 functions that contain at least one log statement.
The statistics of this dataset are shown in Table~\ref{tab:multi_log_dataset}.
This table reports, for each language, the number of functions that contain only one log statement, which is denoted as Single-Log Functions, the number of functions that contain multiple log statements, which is denoted as Multi-Log Functions, and the total number of contained log statements, which is denoted as Total Log Statements.
For the evaluation in Table~\ref{tab:multi_log_llama3}, we use all 200 functions for each language.
We evaluate both Single-Log Functions and Multi-Log Functions.
In practical development environments, the model is not given in advance how many log statements should be included in the target function, and must determine whether to insert a single log statement or multiple log statements.

The purpose of this analysis is not to compare the superiority of specific models, but to confirm the inherent difficulty of a setting in which a function may contain multiple log statements.
We conduct an additional analysis using Llama3, which is one of the representative open-source LLMs consistently used in the main experiments.

In a setting where a function may contain multiple log statements, the number of log statements to be generated is not fixed, so evaluating position prediction only by exact-match accuracy is insufficient.
Following prior work~\cite{Duan2025arXiv}, we use \textit{Precision}, \textit{Recall}, and \textit{F1-score} for log position prediction.
These metrics allow us to evaluate whether the predicted log statements are excessive and may cause system overhead, or insufficient and may miss necessary information.
The metrics related to log levels and messages are calculated only for log statements whose positions are predicted correctly.
\textit{Level Accuracy} represents the proportion of log statements inserted at correct positions whose levels also match.
\textit{Message Accuracy}, \textit{BLEU}, and \textit{ROUGE} represent the agreement and similarity of messages generated for correct positions.
We also use \textit{Function-level} metrics to evaluate the extent to which the model can consistently generate the required log statements for an entire function.
\textit{Function-level Position} represents the proportion of functions for which all log positions in the function are predicted correctly.
\textit{Position+Level} represents the proportion of functions for which both all positions and the levels of the corresponding log statements are predicted correctly.
\textit{Perfect} represents the proportion of functions for which position, level, and message all match for every log statement in the function.
These function-level metrics are necessary because, in practical use, partially correct predictions may still be insufficient if the set of log statements generated for the entire function is inconsistent.

\underline{\textbf{\textit{Results.}}}
Table~\ref{tab:multi_log_llama3} shows the results of Llama3 in this setting.

\textbf{Llama3 can capture some correct log insertion locations, but tends to overpredict unnecessary locations in a setting where a function may contain multiple log statements.}
When averaged across the five languages, \textit{Precision}, \textit{Recall}, and \textit{F1} for position prediction are 22.82\%, 49.69\%, and 31.15\%, respectively.
\textit{Recall} is substantially higher than \textit{Precision}.
This result indicates that Llama3 can capture some correct log insertion locations in a function to some extent, while it also frequently overpredicts unnecessary locations.
The model tends to retrieve correct positions broadly, but has difficulty narrowing the predicted positions appropriately.
For each language, \textit{Precision} remains between 19.11\% and 25.86\%, and the maximum \textit{F1} is only 35.11\%.

\textbf{Generating appropriate log levels and messages becomes even more difficult when multiple log statements must be generated within the same function.}
For level prediction, the average is 51.18\%, while C\# remains at only 23.93\%, indicating substantial variation across languages.
Message generation is even more difficult.
The average \textit{Message Accuracy} is 0.87\%, BLEU-4 is 1.51, and ROUGE-L is 16.90.
\textit{Message Accuracy} is below 2\% for all languages.
This indicates that, when generating multiple log statements appropriately within the same function, it is extremely difficult not only to determine positions but also to generate message content consistently according to the role of each log statement.

\textbf{At the function level, Llama3 rarely generates position, level, and message consistently for the entire function.}
The function-level results clarify the difficulty of this setting.
On average across the five languages, \textit{Function-level Position} is only 4.10\%, \textit{Position+Level} is 1.90\%, and \textit{Perfect} is 0.10\%.
For each language, \textit{Function-level Position} is 7.50\% for JavaScript, 6.00\% for Java, and 5.00\% for TypeScript, whereas it is 0.00\% for Python.
This indicates that matching all log positions in a function exactly is already very difficult.
\textit{Position+Level} is at most 3.00\%, and \textit{Perfect} is 0.00\% for all languages except Java, which reaches only 0.50\%.
Even if some individual log elements can be predicted partially correctly, the model can hardly generate position, level, and message consistently for the entire function at the same time.
This result indicates that the end-to-end generation framework used in the single-log setting does not sufficiently handle the more realistic scenario of inserting multiple log statements.

These results suggest that the difficulty of this setting does not arise simply because the number of prediction targets increases.
It stems from the need to handle, at the same time, the positional relationships among multiple log statements, the division of roles among them, and the consistency of the function as a whole.
In this setting, the model must determine not only where log statements should be inserted, but also how many are necessary, which level should be assigned to each log statement, and how to generate messages that are not redundant with one another.
Conventional generation strategies that assume a single log statement have inherent limitations.
In the future, stepwise generation approaches, which first identify candidate log insertion locations and then generate the level and message for each location, as well as structured approaches that explicitly consider the control flow of the entire function and the relationships among multiple events, are likely to become important.

\subsection{Qualitative Analysis of Generated Log Messages with LLM-as-a-Judge} 
\label{subsec:llm_as_a_judge}

\underline{\textbf{\textit{Motivation.}}}
As demonstrated in the preceding performance analyses, the evaluated log statement generation approaches, including existing end-to-end approaches and LLMs, struggle to achieve high Message Accuracy.
This limitation stems from the strict reliance of the metric on exact string matching against the target log messages; consequently, semantically valid logs are unfairly penalized as incorrect predictions simply because they differ in surface-level phrasing.
Conventional metrics like BLEU and ROUGE primarily measure lexical overlap, and they do not fully capture the semantic equivalence between generated and target log messages.
To fill this gap, we conducted a qualitative analysis with an LLM-as-a-Judge to more accurately assess message quality.
Specifically, our goal was to determine whether a generated log message, despite differing in its exact wording, preserves similar semantics and can serve as a viable alternative to the target log message.

\underline{\textbf{\textit{Approach.}}}
To achieve this, we utilized 15,000 log messages generated by the best-performing approach, UniLog, comparing them directly against the target log messages from the test set.
We employed an LLM-as-a-Judge methodology to assess the semantic equivalence between these pairs.
To systematically classify this semantic similarity, we adopted the taxonomy proposed by~\citet{Mastropaolo2022ICSE}, categorizing each generated instance into one of the following three classes:

\begin{itemize}
    \item \textit{Same Information.} The generated log message is semantically equivalent to the target one, despite differing in its exact lexical expression.
    \item \textit{Meaningful.} The generated log message is contextually coherent and possesses a concrete meaning, but diverges semantically from the intended target message.
    \item \textit{Meaningless.} The generated log message is incomprehensible, incoherent, or entirely irrelevant to the surrounding code context.
\end{itemize}

Based on this taxonomy, we employed GPT-5.2 as an automated annotator to compare the generated log messages against the target ground truth, classifying each generation into one of these three categories.
The prompt used for this automated annotation is shown in Figure~\ref{fig:judge-prompt}.
To empirically validate the LLM-as-a-Judge methodology, we manually verified the automated classifications against a rigorous human evaluation.
Specifically, we randomly sampled 375 instances from the full corpus of 15,000, a size determined to guarantee a 95\% confidence level with a ±5\% margin of error.
The first and second authors served as two independent annotators and classified each instance according to our established taxonomy, achieving an almost perfect inter-annotator agreement with a Cohen's Kappa coefficient of 0.93.
Any isolated labeling discrepancies were subsequently resolved through discussion to establish a definitive consensus.
When comparing these finalized human annotations against the labels assigned by GPT-5.2, we observed a 92.5\% agreement rate.
This robust alignment confirms that our LLM-as-a-Judge methodology serves as a reliable proxy for human judgment in evaluating the semantic similarity between generated and target log messages.
\input{figures/discussion/prompt_figure.tex}

\underline{\textbf{\textit{Results.}}}
\textbf{A semantics-aware evaluation reveals that UniLog generates appropriate log messages more often than exact-match Message Accuracy suggests.}
The analysis revealed that, among the 15,000 log messages generated by UniLog, 5,987 (39.9\%) preserved the same semantic message as the target messages despite surface-level differences, indicating semantic equivalence.
An additional 8,962 (59.8\%) contained meaningful but non-equivalent messages, while the remaining 51 (0.3\%) were classified as meaningless.
Although the Message Accuracy of UniLog was 22.74\%, a semantics-aware evaluation reveals that UniLog generated appropriate messages for 39.9\% of instances, nearly double what exact-match metrics suggest.
These findings indicate that UniLog is capable of generating log messages that are semantically similar to the targets while differing only in surface expression, and that such messages can serve as alternatives.
Consequently, automatic evaluation metrics that rely solely on exact match may substantially underestimate the true capability of UniLog in log message generation.

\input{figures/discussion/llm-as-a-judge_figure.tex}
To further understand the log message generation capability of UniLog, Figure~\ref{fig:llm_as_a_judge} presents representative examples of the three categories: \textit{Same Information}, \textit{Meaningful}, and \textit{Meaningless}.
The first example is a \textit{Same Information} case.
In this example, the synthetic log message (e.g., \texttt{logger.Error("Error whilst closing handle. Type: \{0\}, Handle: \{1\}", requestType, handle, exception);}) can be regarded as semantically equivalent to the developer-written log message (e.g., \texttt{logger.Error("\$\{requestType\} \$\{handle\} error whilst closing handle.", exception);}), although the expression format is different, because both record an error while closing a handle and identify the target associated with that error.
This suggests that UniLog can generate log messages that are practically equivalent in meaning even when their surface expressions differ, based on the given code context.
At the same time, this example shows that Message Accuracy, which is based on exact match, would treat such a semantically appropriate output as incorrect.

The second example is a \textit{Meaningful} case.
In this example, the synthetic log message differs from the developer-written log message (e.g., \texttt{"Unimplemented NID function sceKernelRegisterThreadEventHandler [0x0C106E53]"}) in the identifier at the end (e.g., \texttt{"Unimplemented NID function sceKernelRegisterThreadEventHandler [0xCA145E2E]"}).
The generated message is not semantically equivalent to the target.
It still correctly captures that the log concerns the unimplemented function \texttt{sceKernelRegisterThreadEventHandler}, and it is sufficiently understandable in light of the code context.
In other words, although this generated message does not exactly match the reference, it still carries meaningful information and utility, and is therefore classified as Meaningful.
This example also highlights a limitation of automatic evaluation metrics based on n-grams, such as BLEU and ROUGE.
Because the generated log message shares most of its words and expressions with the target, it is likely to receive a high score under surface-level similarity metrics.
In reality, the difference in the identifier is semantically important, and the content being recorded is not equivalent to the target.
This indicates that n-gram-based evaluation metrics may overestimate surface-level overlap while failing to adequately capture semantic differences.

The third example is a \textit{Meaningless} case.
In this example, the target log message is \texttt{logger.debug("Using maximum allowed span event limit of \%s", maxLimit);}, which records that \texttt{spanLimit} has been restricted to \texttt{maxLimit}.
The synthetic log message is \texttt{\_slicedToArray.\_IGNORE\_ARRAY(vertical\_factors) || (0, \_debug.assert)(false);}, which does not even satisfy the form of a log statement and does not represent the content that should be recorded in the given code context.
This generated output is semantically inappropriate and is classified as Meaningless.
Such an example indicates that, although UniLog can generate contextually appropriate log messages in many cases, there are still cases in which it produces outputs that deviate substantially in terms of contextual understanding and output form.

Overall, these findings underscore that evaluating log message generation requires a paradigm shift beyond surface-level matching metrics like Message Accuracy, BLEU, and ROUGE.
Specifically, while strict exact-match criteria (Message Accuracy) consistently underestimate model capabilities by penalizing \textit{Same Information} outputs that merely differ in phrasing, $n$-gram-based metrics (BLEU and ROUGE) risk artificially inflating scores by rewarding lexical overlap in \textit{Meaningful} but semantically non-equivalent messages.
To accurately gauge the utility of advanced log generation approaches like UniLog, future evaluations should complement conventional automated metrics with semantic assessments, such as LLM-as-a-Judge frameworks and human validation.

%% file: figures/discussion/cross_language_perfect_heatmap_figure.tex
\begin{figure}[t]
  \centering
  \includegraphics[width=0.5\textwidth]{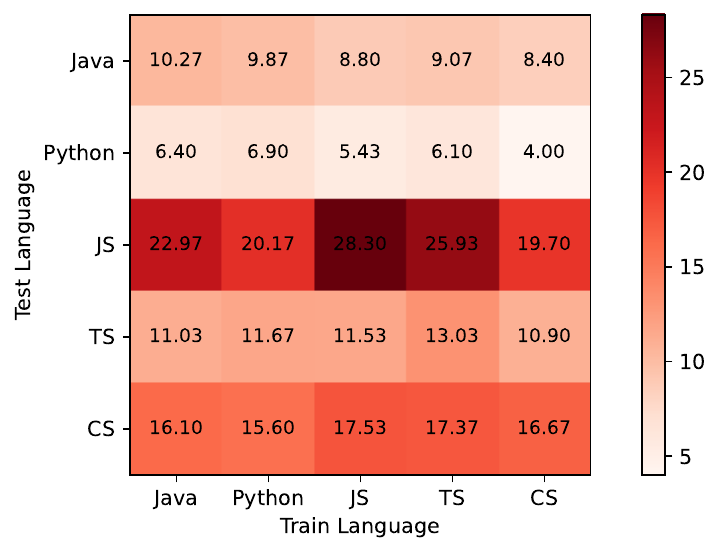}
  \caption{Heatmap of All Accuracy for cross-language log statement generation.}
  \label{fig:discussion_cross_language_perfect}
\end{figure}

%% file: table/discussion/target_wise.tex
\begin{table}[t]
\centering
\caption{Target-wise comparison between in-language and cross-language LoRA models based on All Accuracy.}
\label{tab:cross_language_target_summary}
\begin{tabular}{lcccc}
\toprule
\textbf{Target} & \textbf{In-lang} & \textbf{Cross-lang Avg.} & \textbf{Diff} & \textbf{Best Source} \\
\midrule
Java        & 10.73 & 8.84  & 1.89 & Python \\
Python      & 6.90  & 5.48  & 1.42 & Java \\
JavaScript  & 28.30 & 22.19 & 6.11 & TypeScript \\
TypeScript  & 13.03 & 11.28 & 1.75 & Python \\
C\#         & 16.67 & 16.65 & 0.02 & JavaScript \\
\bottomrule
\end{tabular}
\end{table}

%% file: table/discussion/multi-dataset.tex
\begin{table}[t]
\centering
\caption{Statistics of the multi-log-statement dataset}
\label{tab:multi_log_dataset}
\begin{tabular}{lccc}
\toprule
\textbf{Language} & \textbf{Single-Log Functions} & \textbf{Multi-Log Functions} & \textbf{Total Log Statements} \\
\midrule
Java       & 134 & 66  & 342 \\
Python     & 97  & 103 & 432 \\
JavaScript & 117 & 83  & 345 \\
TypeScript & 109 & 91  & 373 \\
C\#        & 123 & 77  & 361 \\
\midrule
Total      & 580 & 420 & 1,853 \\
\bottomrule
\end{tabular}
\end{table}

%% file: table/discussion/multi_log_results.tex
\begin{table}[t]
\centering
\caption{Overall performance of Llama3 on the multi-log-statement setting}
\label{tab:multi_log_llama3}
\resizebox{\textwidth}{!}{
\begin{tabular}{lcccccccccc}
\toprule
\multirow{2}{*}{\textbf{Language}} & \multicolumn{3}{c}{\textbf{Position}} & \textbf{Level} & \multicolumn{3}{c}{\textbf{Message}} & \multicolumn{3}{c}{\textbf{Function}} \\
\cmidrule(lr){2-4} \cmidrule(lr){5-5} \cmidrule(lr){6-8} \cmidrule(lr){9-11}
& \textbf{Precision} & \textbf{Recall} & \textbf{F1}
& \textbf{Acc.}
& \textbf{Acc.} & \textbf{BLEU-4} & \textbf{ROUGE-L}
& \textbf{Position} & \textbf{Position+Level} & \textbf{Perfect} \\
\midrule
Java       & 19.11 & 54.09 & 28.24 & 63.78 & 0.54 & 0.19 & 20.26 & 6.00 & 3.00 & 0.50 \\
Python     & 22.21 & 46.99 & 30.16 & 62.56 & 0.99 & 2.11 & 19.12 & 0.00 & 0.00 & 0.00 \\
JavaScript & 24.81 & 47.54 & 32.60 & 51.22 & 1.83 & 2.46 & 17.12 & 7.50 & 3.00 & 0.00 \\
TypeScript & 25.86 & 54.69 & 35.11 & 54.41 & 0.98 & 2.54 & 20.29 & 5.00 & 3.00 & 0.00 \\
C\#        & 22.09 & 45.15 & 29.66 & 23.93 & 0.00 & 0.27 & 7.73  & 2.00 & 0.50 & 0.00 \\
\midrule
Average    & 22.82 & 49.69 & 31.15 & 51.18 & 0.87 & 1.51 & 16.90 & 4.10 & 1.90 & 0.10 \\
\bottomrule
\end{tabular}
}
\end{table}

%% file: figures/discussion/prompt_figure.tex
\begin{figure}[t]
  \centering
  \includegraphics[width=0.9\textwidth]{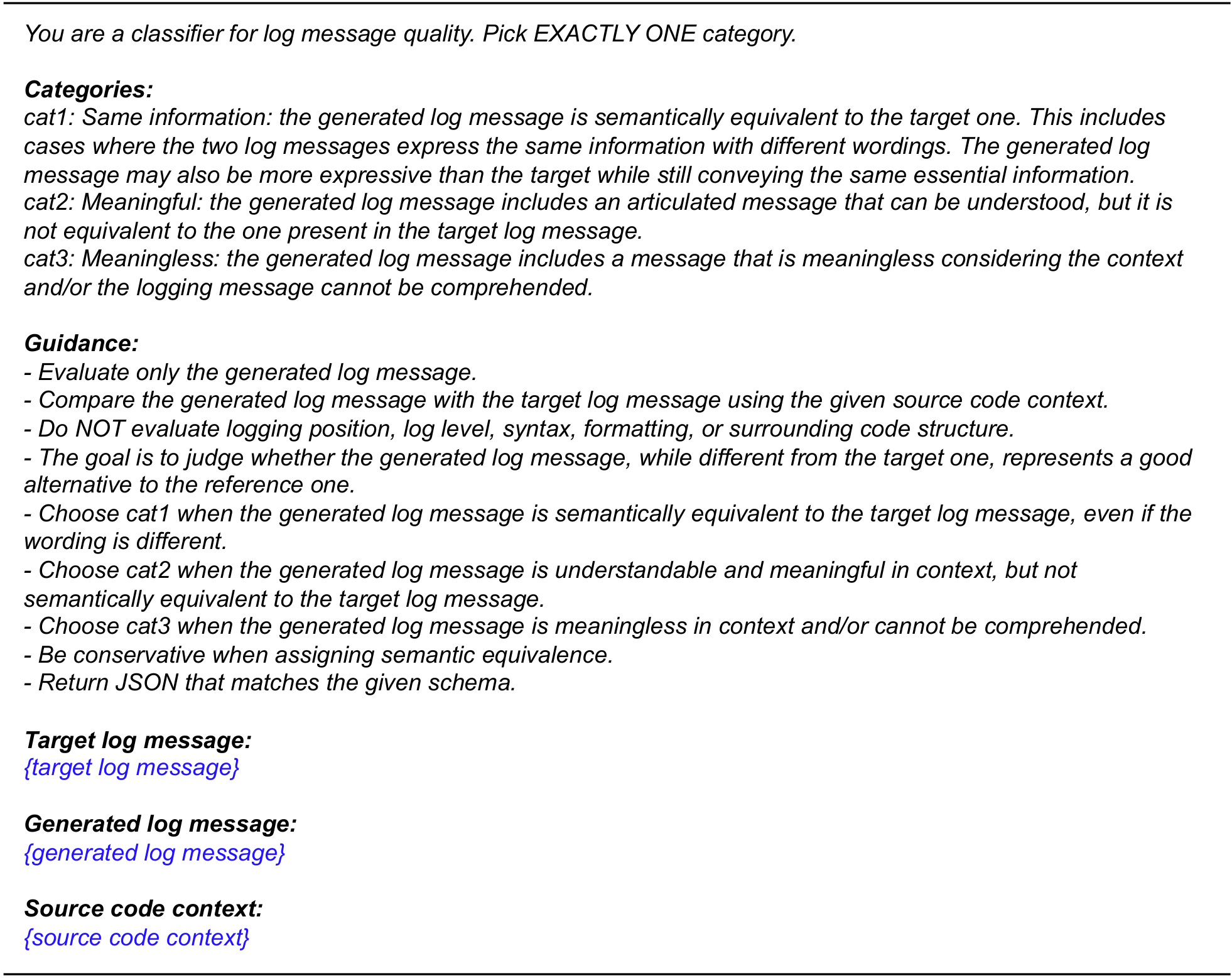}
  \caption{Prompt used to classify generated log messages in the LLM-as-a-Judge evaluation.}
  \label{fig:judge-prompt}
\end{figure}

%% file: figures/discussion/llm-as-a-judge_figure.tex
\begin{figure}[t]
  \centering
  \includegraphics[width=0.9\textwidth]{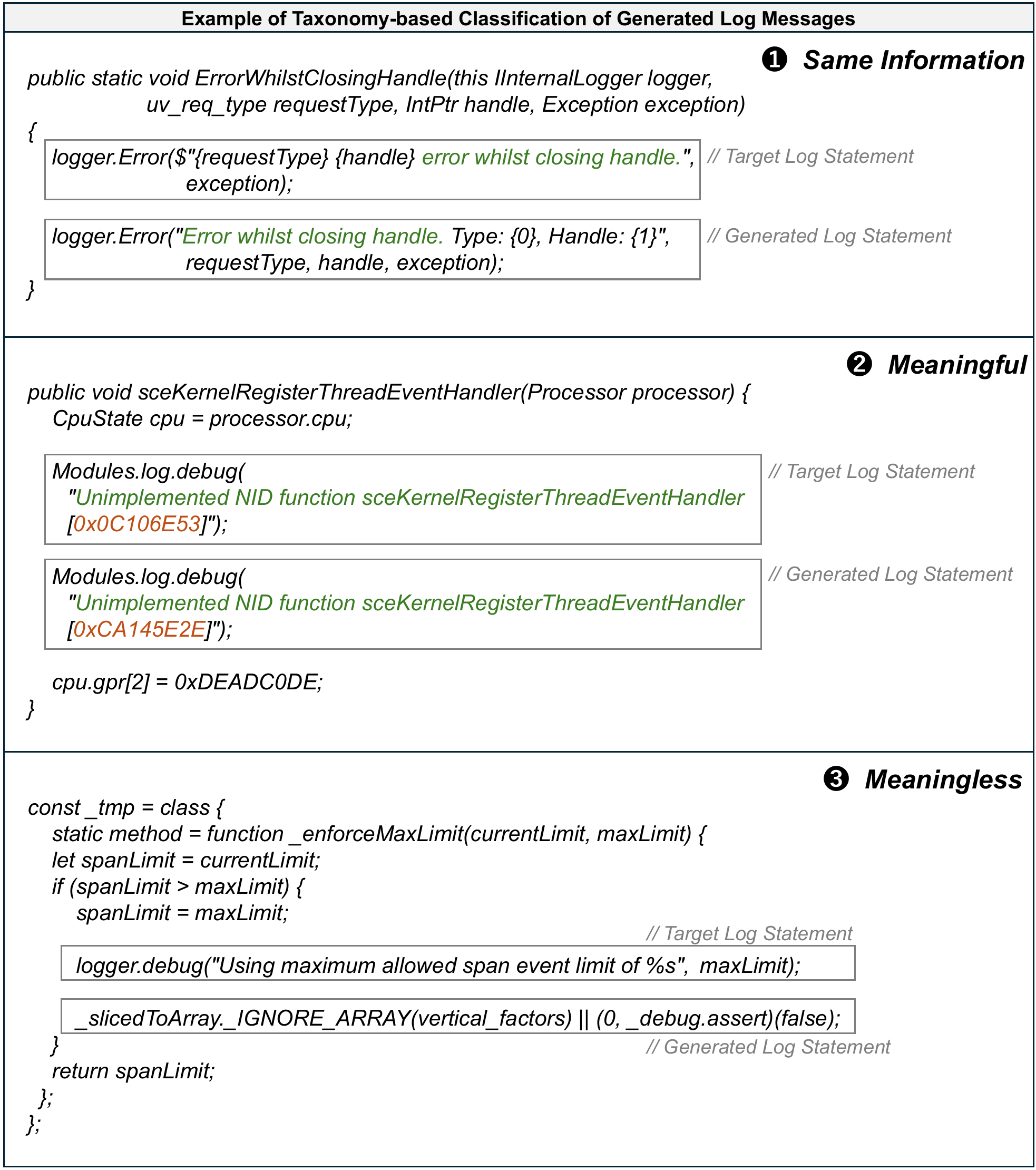}
  \caption{Examples of Log Message Generated by UniLog.}
  \label{fig:llm_as_a_judge}
\end{figure}

%% file: section/threats.tex
\section{THREATS} \label{sec:threats}
In this section, we discuss threats to validity in this study.

\subsection{Construct Validity}
In this study, we selected target projects by applying strict criteria, such as requiring at least 500 commits and at least 10 contributors.
The purpose was to exclude toy projects and, by extension, code that is likely to be of low quality.
This is consistent with prior logging-related studies, including LANCE~\cite{Mastropaolo2022ICSE}, UniLog~\cite{Xu2024ICSE}, and FastLog~\cite{Xie2024ISSTA}.
Low-quality code instances may still be included in the training set or the test set.
In addition, the category classification of log statement locations was automated by using GPT-5.2, but this process may include misclassifications.
To address this threat, we conducted manual labeling by two authors on 375 randomly sampled representative instances, reported Cohen's kappa to show the agreement between the two authors, and further reported the agreement between the GPT-5.2 classification results and the human labels in order to assess the reliability of this process.

\subsection{Internal Validity}
One potential threat is whether LANCE and UniLog were faithfully reproduced.
Because no public implementation of UniLog is available, we reimplemented it based on the original paper~\cite{Xu2024ICSE}, but inconsistencies may have arisen.
To mitigate this threat, we included the implementation in the replication package so that future researchers can verify it more easily.
Although an implementation of LANCE is publicly available, the original implementation no longer worked because the Google Colab environment had been updated, and we therefore modified it so that it could run in our environment.
Because of this modification, the implementation of LANCE may not be perfectly consistent with the original paper~\cite{Mastropaolo2022ICSE}.
To mitigate this threat, we used the dataset from the original LANCE paper and confirmed that the modified implementation achieved performance comparable to that reported in the original paper, within 3 percentage points.
Another potential threat is data leakage.
For example, UniLog uses GPT-4.1 mini, which was pre-trained on data up to April 2025, and therefore some of the evaluated data may have been included in the pre-training corpus, which may have inflated the performance.

\subsection{External Validity}
The main potential threat concerns the choice of LLMs used in this study.
We selected five LLMs, Llama3, Qwen2.5-Coder, Mistral, GPT-4.1 mini, and DeepSeek-V3.
Although these models have also been used in prior work~\cite{Zhong2025TOSEM,Kazuki2025ESEM,shu2025EMSE,Acharya2025EASE,zan2025NeurIPS,Wang2025SC}, it remains unclear whether similar tendencies would be observed if other LLMs were used.





%% file: section/conclusion.tex
\section{Conclusion} \label{sec:conclusion}
In this study, we systematically evaluated log statement generation approaches, including existing end-to-end approaches and LLMs, and analyzed the impact of training strategies in multilingual environments (i.e., Java, Python, JavaScript, TypeScript, and C\#).
In the unified performance comparison, UniLog achieved the best overall performance among all evaluated approaches, and it can relatively stably align the three elements (i.e., insertion location, log level, and message content) even in multilingual environments.
The analysis of training strategies confirmed that language-specific training for each single language tends to achieve higher performance than multilingual settings that train multiple languages simultaneously.
In addition, the training strategy of UniLog achieved higher performance than a LoRA model trained with 120k instances, although it uses only 500 validation queries for warmup.
This result indicates that, for log statement generation, the overall design of the approach, including the training procedure and prompt design, has a strong impact on performance, rather than simply increasing the amount of training data.
We also confirmed that the difficulty of log statement generation differs substantially across programming languages.
Our detailed analysis revealed that language-specific logging characteristics, including insertion-location distributions, log-level distributions, and message diversity, are strongly related to prediction difficulty.
These results indicate that, in multilingual log statement generation, it is not sufficient simply to increase model size or training data size, and that approach design that takes into account language-specific logging practices and insertion-category characteristics is important.

This study clarifies, under the same conditions, the capabilities of existing end-to-end approaches and LLMs for multilingual log statement generation, and provides important insights into which factors become bottlenecks for improving performance.
In particular, overall performance is still strongly constrained by All Accuracy, which requires simultaneous correctness of Position, Level, and Message.
Future improvements require model designs and training strategies that consider the characteristics of programming languages and language-specific logging practices.
As promising future research directions, in addition to designing training approaches that leverage logging knowledge shared across languages, evaluation under settings that are closer to real deployment, such as multi-log-statement insertion and cross-language transfer, is also promising.
We expect this study to provide a foundation for the development of robust log statement generation approaches that are practically usable in multilingual environments.




%% file: section/data_availability.tex
\section{Data Availability} \label{sec:data_availability}
Our replication package can be accessed at \url{https://doi.org/10.5281/zenodo.20279312}.